\documentclass{aa}  

\usepackage{graphicx}
\usepackage{txfonts}
\usepackage{tikz}
\usetikzlibrary{calc}
\usetikzlibrary{positioning}
\usetikzlibrary{shapes.geometric}

\begin{document}

   \title{AutoSourceID-FeatureExtractor}

   \subtitle{Optical image analysis using a two-step mean variance estimation network for feature estimation and uncertainty characterisation}

   \author{F. Stoppa \inst{1,6}
        \and
        R. Ruiz de Austri \inst{2}
        \and
        P. Vreeswijk \inst{1}
        \and
        S. Bhattacharyya \inst{5}
        \and
        S. Caron \inst{3,4}
        \and
        S. Bloemen \inst{1}
        \and 
        G. Zaharijas \inst{5,9}
        \and
        G. Principe \inst{10,11}
        \and
        V. Vodeb \inst{5}
        \and
        P.J. Groot \inst{1,12,13}
        \and
        E. Cator\inst{6}
        \and
        G. Nelemans \inst{1,7,8}
      }

   \institute{
            Department of Astrophysics/IMAPP, Radboud University, PO Box 9010,
            6500 GL Nijmegen, The Netherlands
            \and
            Instituto de Física Corpuscular, IFIC-UV/CSIC, Valencia, Spain
            \and
            High Energy Physics/IMAPP, Radboud University, PO Box 9010,
            6500 GL Nijmegen, The Netherlands
            \and
            Nikhef, Science Park 105, 1098 XG Amsterdam, the Netherlands
            \and
            Center for Astrophysics and Cosmology, University of Nova Gorica, Vipavska 13, SI-5000 Nova Gorica, Slovenia
            \and
            Department of Mathematics/IMAPP, Radboud University, PO Box 9010, 6500 GL Nijmegen, The Netherlands
            \and
            SRON, Netherlands Institute for Space Research, Sorbonnelaan 2, NL-3584 CA Utrecht, The Netherlands
            \and
            Institute of Astronomy, KU Leuven, Celestijnenlaan 200D, B-3001 Leuven, Belgium
            \and 
            Institute for Fundamental Physics of the Universe, Via Beirut 2, 34151 Trieste, Italy
            \and
            Dipartimento di Fisica, Universit\'a di Trieste, I-34127 Trieste, Italy
            \and
            Istituto Nazionale di Fisica Nucleare, Sezione di Trieste, I-34127 Trieste, Italy
            \and
            Department of Astronomy and Inter-University Institute for Data Intensive Astronomy, University of Cape Town, Private Bag X3, Rondebosch, 7701, South Africa
            \and
            South African Astronomical Observatory, P.O. Box 9, Observatory, 7935, South Africa
            }

   \date{\today}

\abstract
{}
{In astronomy, machine learning has been successful in various tasks such as source localisation, classification, anomaly detection, and segmentation. However, feature regression remains an area with room for improvement. We aim to design a network that can accurately estimate sources' features and their uncertainties from single-band image cutouts, given the approximated locations of the sources provided by the previously developed code AutoSourceID-Light (ASID-L) or other external catalogues. This work serves as a proof of concept, showing the potential of machine learning in estimating astronomical features when trained on meticulously crafted synthetic images and subsequently applied to real astronomical data.}
{The algorithm presented here, AutoSourceID-FeatureExtractor (ASID-FE), uses single-band cutouts of 32x32 pixels around the localised sources to estimate flux, sub-pixel centre coordinates, and their uncertainties. ASID-FE employs a two-step mean variance estimation (TS-MVE) approach to first estimate the features and then their uncertainties without the need for additional information, for example the point spread function (PSF). For this proof of concept, we generated a synthetic dataset comprising only point sources directly derived from real images, ensuring a controlled yet authentic testing environment.}
{We show that ASID-FE, trained on synthetic images derived from the MeerLICHT telescope, can predict more accurate features with respect to similar codes such as SourceExtractor and that the two-step method can estimate well-calibrated uncertainties that are better behaved compared to similar methods that use deep ensembles of simple MVE networks. Finally, we evaluate the model on real images from the MeerLICHT telescope and the Zwicky Transient Facility (ZTF) to test its transfer learning abilities.}
{}


\maketitle
%


\section{Introduction}
\label{sec: Intro}

Optical sky surveys have a significant impact on various scientific fields, such as astrophysics, cosmology, and planetary science. 
These surveys collect data on the positions, luminosities, and additional attributes of celestial objects, such as stars, galaxies, and quasars, and have contributed greatly to the scientific community's knowledge and understanding of the universe.

Shortly, the development of multiple large-scale optical survey telescopes will substantially enhance our capacity to investigate the cosmos and expand our understanding of its properties and evolution. A prime example is the Vera C. Rubin Observatory \citep{Ivezic2019}, which is presently under construction in Chile. Equipped with an 8.4-metre telescope engineered for conducting an extensive sky survey spanning over 20,000 square degrees, it will deliver crucial data on billions of celestial objects. Furthermore, the adoption of large-format complementary metal-oxide semiconductor (CMOS) detectors in astronomy is expected to increase data acquisition speeds by two orders of magnitude, leading to a substantial increase in data volume.

Due to the imminent surge of data, traditional image processing methods will face escalating challenges. Consequently, recent years have witnessed a transition towards using machine learning techniques for data analysis, with the ultimate objective of real-time processing. Specifically, within the specialised field of astronomical image processing, convolutional neural networks (CNNs, \citealt{Lecun1998}) have proven remarkably effective, excelling in a variety of applications such as galaxy classification based on morphological characteristics \citep{Vavilova2022}, exoplanet detection \citep{Cuellar2022}, image reconstruction from noise-corrupted or incomplete data \citep{Flamary2016}, detection and classification of point sources from gamma-ray data \citep{Panes2021}, and photometric redshift estimation \citep{Mu2020, Schuldt2021}.

Nevertheless, while there have been some applications of CNNs for photometric redshift estimation in astronomy \citep{Hoyle2016, D'Isanto2018, Pasquet2019}, their employment for feature regression tasks, such as estimating flux and other source properties, remains largely unexplored. In this context, feature regression involves predicting continuous numerical output values based on features extracted from images, requiring the network to manage the complexity of image data as input while simultaneously performing a regression task.
Several factors may account for this underutilisation. First, the complexity of astronomical tasks could be a significant barrier. Second, the lack of adequately labelled data for training these networks may limit their use. Lastly, the novelty of applying CNNs in this context might lead to hesitance within the astronomical community. 
Despite these challenges, they represent unique opportunities for groundbreaking work in this promising intersection of deep learning and astronomy. 

A shared concern in both domains is the reliable estimation of uncertainties, which is essential for ensuring the robustness of astronomical findings. While deep neural networks (DNNs) have achieved remarkable performance across various applications \citep{LeCun2015, Schmidhuber2015, Goodfellow2016}, their black-box nature often hinders their ability to quantify predictive uncertainties effectively. Beyond merely predicting an expected value, it is crucial to gauge and understand the associated uncertainty. Such an approach, as highlighted by many others \citep{Rasmussen2004, Kiureghian2009, Ghahramani2015, kendall2017, Smith2018, Wilson2020}, not only enhances the reliability and confidence of predictions but also facilitates more informed decision-making. This reliable estimation of uncertainties will also be a central theme of this paper.

Another pivotal aspect in the realm of astronomical data processing is the use of synthetic astronomical image generators. As the intersection of machine learning and astronomy continues to evolve, it is essential to recognise the burgeoning role of these sophisticated tools.
Tools such as Pyxel \citep{Arko2022} and ScopeSim \citep{Leschinski2020} are leading this advancement, striving to produce images with an unparalleled level of detail that encapsulate everything from subtle celestial features to a broad spectrum of observational artefacts. While the astronomical landscape may not be dominated by these synthetic images, their increasing fidelity makes them invaluable assets for machine learning training. By utilising these high-quality synthetic images, we can ensure that machine learning models are trained on true features and are adeptly prepared to handle the nuances of real-world astronomical data. This paper is a testament to our commitment to this progressive approach. It introduces one of the core data processing techniques that we intend to integrate into our envisioned comprehensive pipeline. This pipeline is designed to harness the full potential of machine learning tools, from source detection and feature regression to transient identification, with the ultimate goal of establishing the first fully automated machine learning-driven telescope pipeline.

Building on this vision and recognising the need for practical tools and methodologies, for this study, we tackled two objectives. First, we constructed a network capable of predicting, from image cutouts of sources detected with AutoSourceID-Light (ASID-L, \citealp{Stoppa2022}) or through other tools, their flux, sub-pixel centre coordinates, and corresponding uncertainties. Second, we aimed to enhance the field of uncertainty estimation in machine learning by implementing what we call a two-step mean variance estimation (TS-MVE) network.
In Section \ref{sec: Method}, we demonstrate the concept on a synthetic dataset and highlight its improvements from an astronomical perspective and an uncertainty characterisation standpoint. Subsequently, we compare its outcomes with SourceExtractor \citep{Bertin1996} and another DNN for regression that adopts an ensemble methodology \citep{Lakshminarayanan2017}.
Lastly, as explained in Section \ref{sec: real application}, we applied the trained model to a real set of images from the MeerLICHT \citep{Bloemen2016} and the Zwicky Transient Facility (ZTF, \citealp{Bellm2019}) telescopes to evaluate its transfer learning capabilities.

\noindent
The code presented here\footnote{\url{https://github.com/FiorenSt/AutoSourceID-FeatureExtractor}} is the third deep learning algorithm developed within the context of MeerLICHT/BlackGEM \citep{Bloemen2016,groot2022}, following MeerCRAB \citep{Hosenie2021}, an algorithm employed to classify real and bogus transients in optical images, and ASID-L \citep{Stoppa2022b}, an algorithm designed for rapid source localisation in optical images.

\section{Data}
\label{sec: Data}

An appropriate experimental protocol must be established to assess the efficacy of the proposed two-step network for feature regression and validate the core concept of the proposed method. One vital component of this protocol is employing both synthetic and real MeerLICHT telescope image datasets to train and evaluate the network performance. While the real dataset provides real-world scenarios and helps assess the network's performance in practical applications, the synthetic dataset is necessary to train the network on true, exact features.

\subsection{MeerLICHT telescope}
\label{sec: MeerLICHT_telescope}

The 65 cm optical MeerLICHT telescope has a $2.7$ square degree field of view and a 10.5k $\times$ 10.5k pixel CCD. Its primary objective is to track the pointings of the MeerKAT radio telescope \citep{Jonas2016}, facilitating the concurrent detection of transients in both radio and optical wavelengths. The telescope employs the Sloan Digital Sky Survey (SDSS) $ugriz$ filter set and an extra wide $g+r$ filter called $q$ (440-720nm bandpass). Images captured are promptly sent to the IDIA/ilifu facility, where BlackBOX image processing software \footnote{\url{https://github.com/pmvreeswijk/BlackBOX}}, Vreeswijk et al., in prep) handles the images using standard methods. This includes source detection (currently with SourceExtractor, \citealp{Bertin1996}), astrometric and photometric calibration, determination of position-dependent image point spread function (PSF), image subtraction, and transient detection.

\subsection{Synthetic dataset}
\label{sec: Synthetic_dataset}

Using synthetic datasets for model training and evaluation before testing on real-world data is a prevalent practice in deep learning and computer vision; it offers greater control over the problem, diminishes the impact of unforeseen variations in real-world data, and enables more efficient use of computational resources.
Real data quality, in fact, can be impacted by various factors, such as atmospheric conditions, telescope optics, and camera noise. To address these constraints and to have exact labelled features, researchers have investigated using synthetic data for training machine learning models \citep{Tremblay2018}. 

The synthetic dataset built for this paper is designed to comprise a substantial number of image samples, approximately 500 full-field MeerLICHT images, each associated with a set of target feature values. The synthetic images are built using the astrometric calibration, the characterisation of the PSF and the photometric calibration of actual MeerLICHT images. These products are kept for each image following the original image processing. The astrometric calibration is accurate to about 0.03" and is inferred using Astrometry.net \citep{astrometry.net} with Gaia DR2 \citep{Gaia2018} index files. The position-dependent PSF for each image is determined using PSFEx \citep{PSFEx}, which uses hundreds of selected stars across the image to fit their profile, where the PSF is allowed to vary as a function of X- and Y-position with a 2nd-order polynomial. From the PSFEx products, a pixel map of the PSF at each position in the image can be created, with a radius of 5 times the average image full-width at half maximum (FWHM) inferred from the stellar profiles (the PSF image is actually square, but the pixels at distances larger than a radius of 5 times the FWHM are set to zero). The photometric calibration is based on the photometry of stars down to $\sim17th$ magnitude from a combination of surveys (Gaia DR2, SDSS, Pan-STARRS, SkyMapper, GALEX and 2MASS) for which stellar templates are fit. The MeerLICHT magnitudes are inferred from the best-fit template, taking into account the typical atmospheric conditions at Sutherland and the wavelength-dependent transmission of the telescope, including the mirror reflectivity, filter transmission and CCD sensitivity. This provides an all-sky calibration catalogue that is used to calibrate each MeerLICHT image, which typically contains hundreds of calibration stars, to an accuracy of $\sim0.02$ mag in the q-band. When inferring the instrumental flux of the sources to be compared to the calibrated fluxes, the source flux is weighted with the PSF at the source position; this optimal flux determination closely follows the method described in \citet{Horne1986}.

Using these three components from an actual MeerLICHT image, we project all relevant sources from the Gaia catalogue onto a blank image with the same size as a reduced MeerLICHT image ($10560 \times 10560$ pixels). In this projection, the Gaia G-band magnitude is converted to flux (unit: electrons/s) using the image q-band zero-point, taking into account the actual airmass of the image and the typical q-band extinction coefficient. This flux determines the amplitude of the PSF pixel map inferred at each particular location so that the total flux (now in units of electrons with the image exposure time taken into account) is the volume below the PSF surface. Although the Gaia G-band filter is very different from the MeerLICHT q-band (despite having similar central wavelengths), this is not important as the resulting flux is simply adopted as the true flux of the object. Besides the Gaia sources, we also added the sky background image and its standard deviation image (inferred on sub-images of size $60 \times 60$ pixels, with the detected sources masked out) of the original image; similar to the PSF characterisation, these are pipeline products that are kept for all MeerLICHT images. In this way, a sample of 500 synthetic images has been built from actual MeerLICHT images, reflecting the actual observing conditions under which the images were taken, with a wide range of image quality and limiting magnitudes.

\begin{figure}
\includegraphics[scale=0.42]{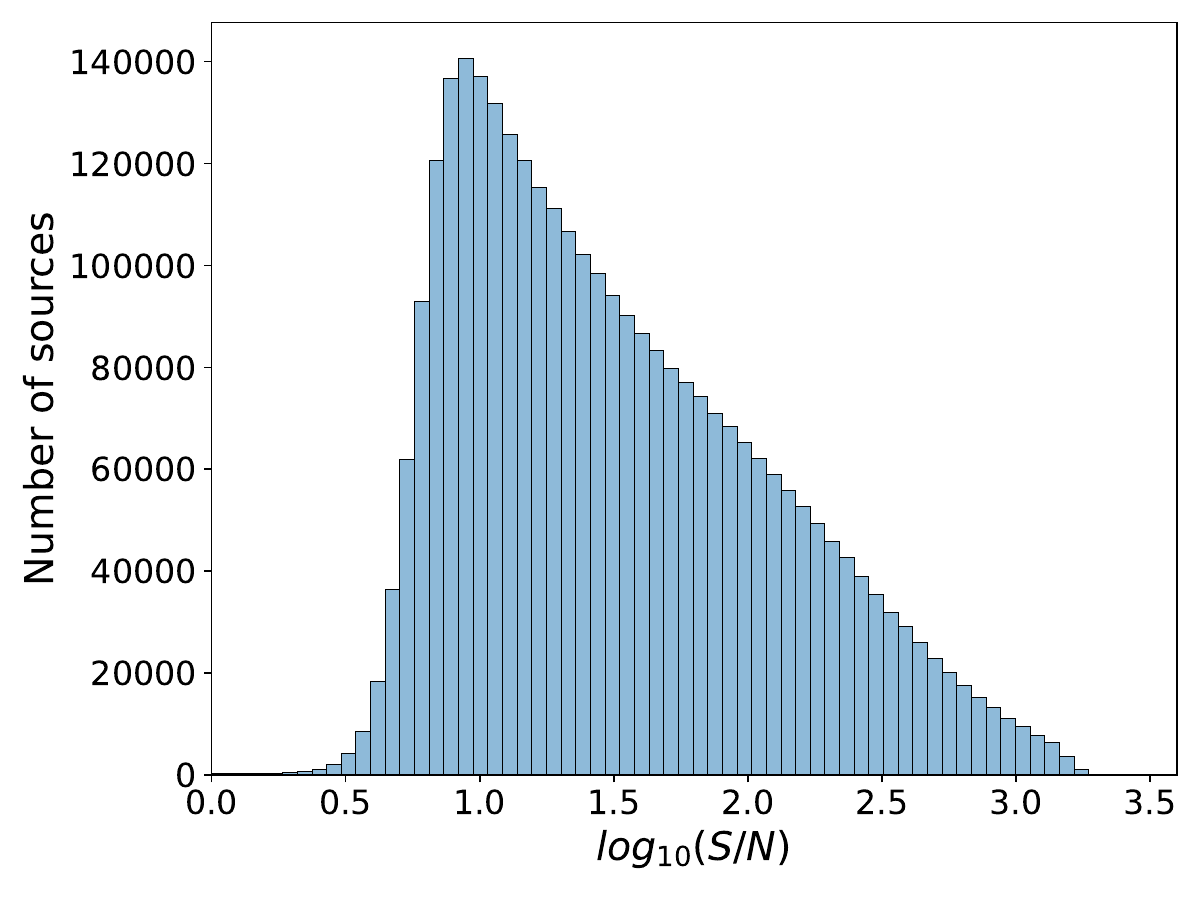}
\caption{Number of sources as a function of S/N. This plot showcases the range of S/N present in our synthetic dataset, emphasising the variety of conditions under which the network will need to be trained and evaluated. A significant portion of the sources exhibit low S/N, underscoring the challenges the network will face in accurately detecting and analysing these sources.}
\label{fig: Counts}
\end{figure}

In each synthetic image, sources are identified using both ASID-L and SourceExtractor. The mutually identified sources are then matched with their true positions in the synthetic images to generate single-band cutouts. This process results in approximately 3.5 million cutouts, each measuring 34x34 pixels.
To further elucidate the characteristics of our synthetic dataset, we have plotted the number of sources as a function of the signal-to-noise ratio (S/N). Training a network to accurately detect and analyse such low S/N sources presents a considerable challenge, given the inherent difficulty in distinguishing these faint sources from the background. Fig. \ref{fig: Counts} provides insights into the distribution of sources across different S/N values, highlighting the diversity and richness of our dataset.

\begin{figure}
\includegraphics[scale=0.45]{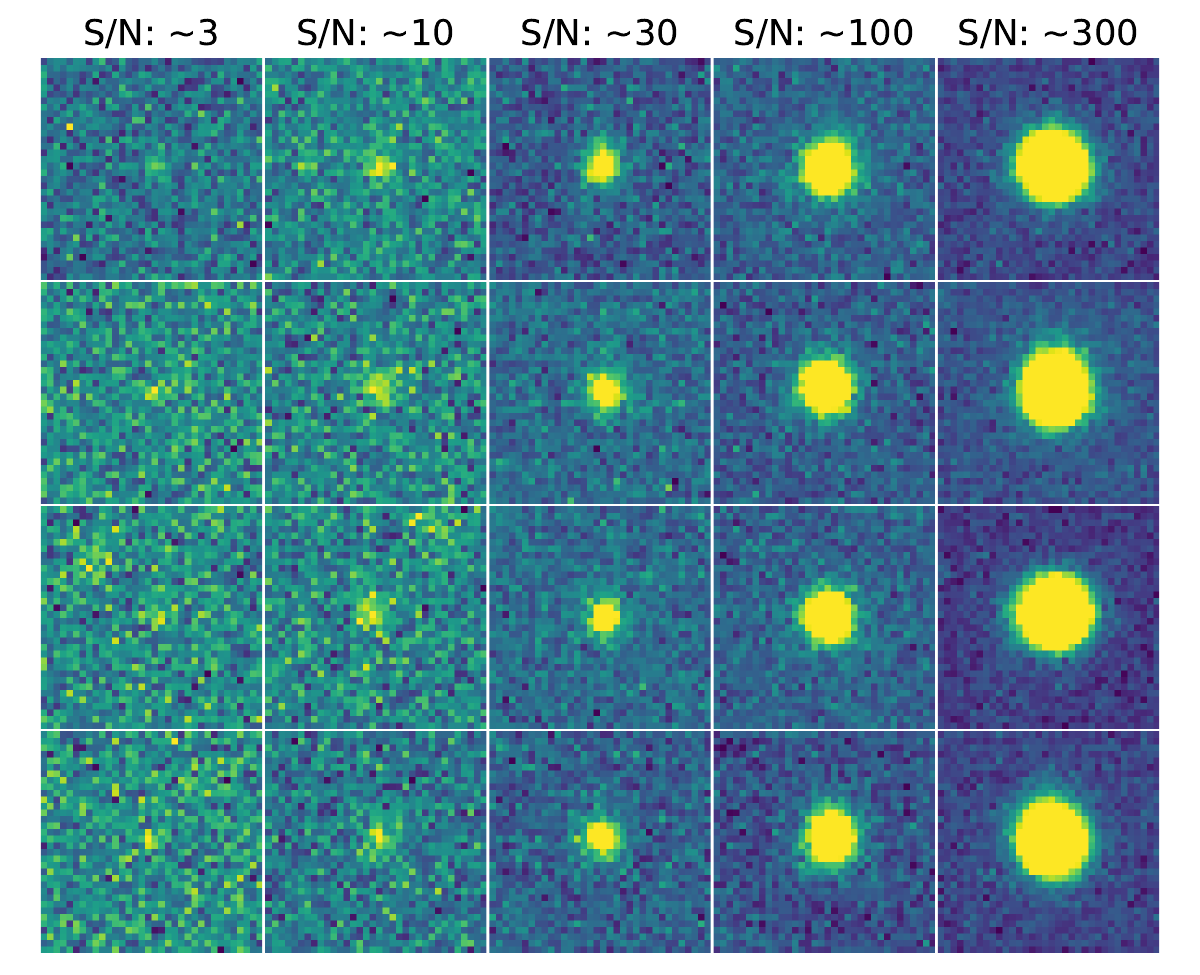}
\caption{Random selection of cutouts from the synthetic exposures showcasing the variety and quality of the simulated data. This visual representation aids in understanding the conditions and challenges the network is designed to handle.}
\label{fig: Cutouts}
\end{figure}

\noindent
To provide a clearer visual understanding of the cutouts and a glimpse into the conditions under which the regressor operates, in Fig. \ref{fig: Cutouts}, we show a random selection of cutouts at different S/N. 

The dataset is then split into $60\%$ training, $20\%$ test, and $20\%$ validation. The true centre in pixel coordinates and flux were stored and incorporated into the final training set for each source. Even though the stored cutouts have dimensions of 34x34 pixels, the training process is carried out on 32x32 pixel images. Having additional pixels at the edges of the cutouts allows us to shift the location of the sources within the image dynamically during training, serving a dual purpose: providing a more varied dataset and simulating suboptimal source localisation by a prior tool. This approach enhances the training dataset and prepares the model to handle situations where source localisation is less than ideal, thereby improving its overall performance and adaptability in various scenarios. 

\noindent
This controlled environment is ideal for training our network and assessing its performance with alternative architectures, such as deep ensemble methods and widely used algorithms like SourceExtractor.

\subsection{Real dataset}
\label{sec: Real_dataset}

While synthetic datasets have advantages, they inevitably lack the full range of real-world complexities. Therefore, it is crucial to validate the performance of the trained models on a real dataset as well. The real dataset used for this study comprises the same set of 500 full-field images captured by the MeerLICHT telescope, identical to those used to create the synthetic dataset. Each image in the real dataset matches the dimensions of those in the synthetic dataset, being 10.5k $\times$ 10.5k pixels. Moreover, the creation of cutouts from the real dataset follows the same methodology employed for the synthetic images, as detailed in the previous section.

The processing of this dataset, however, is quite different. Source extraction is performed using SourceExtractor \citep{Bertin1996}, a widely used astronomical source detection tool that identifies and measures the properties of sources in the given images. In contrast to the synthetic dataset, where true feature values are known precisely, this software provides estimated values for various features such as flux, position, and shape parameters based on the real image data. Consequently, these values can be impacted by various factors such as the background noise level, the complexity of the object itself, and even the subtle intricacies of the software's algorithms. As a result, unlike the synthetic data, the features obtained from this real data do not have associated true values. However, the real dataset is invaluable for assessing the network's performance in real-world scenarios, and it provides an essential test of the generalisation ability of the trained models.

\section{Method}
\label{sec: Method}

In astronomy, various methods exist to infer a star's flux in an optical image. The most prevalent techniques include aperture photometry, which entails positioning a circular aperture around the star and measuring the flux within that aperture \citep{Golay1974}, and profile-fitting photometry, which involves fitting a model of the PSF of the telescope to the star's image \citep{Heasley1999}. However, these methods exhibit limitations, particularly when handling crowded fields or images with a low S/N. 

This paper proposes a machine learning approach to estimate the flux and sub-pixel centre position, $x$ and $y$, of the optical sources, along with associated uncertainties, without using PSF information.
The quantities of interest in the problem, flux, x, and y, can be collectively represented as $\beta=\{x, y, flux \}$. These true quantities are related to the estimated values, $\hat{\beta}$, through a model of the form:
$$\beta = \hat{\beta} + \epsilon,$$

\noindent
where $\epsilon$ denotes the residuals or errors in the estimate, which are assumed to follow a Gaussian distribution with a mean of zero and a covariance matrix $\Sigma$, that is, $\epsilon \sim N(0,\Sigma)$. Assuming Gaussian distributed residuals allows us to estimate not only the quantities of interest, $\hat{\beta}$, but also the uncertainties in these estimates, namely the covariance matrix $\hat{\Sigma}$ of $\epsilon$. 

This Gaussian residual assumption is particularly valid when our synthetic images closely emulate real images in every nuance, including the presence of artefacts. In such cases, the neural network is trained to account for these intricacies, leading to residuals that predominantly follow a Gaussian distribution. This is because the network would have learned to adjust for these artefacts during the training process, and any deviations from the predictions would be due to random noise, which is often Gaussian in nature.
However, the efficacy of our model heavily hinges on the fidelity of the synthetic images to real-world images. If the synthetic images fall short of capturing the intricate complexities inherent in real images, it can lead to residuals that stray from a Gaussian distribution. Specifically, issues such as contamination from image artefacts and confusion noise at low source density can introduce pronounced non-Gaussian noise contributions, undermining the effectiveness of our proposed model.
Beyond the fidelity of synthetic images, intrinsic data complexities can lead to non-Gaussian residuals. For instance, in spectroscopic multiple-star systems, photometric parameters may not be directly inferred from the data and instead need to be constrained by prior knowledge, such as previous observational data, theoretical models, or known astrophysical constraints. Similarly, for complex sources such as galaxies or areas with saturation, the interpretation of features can introduce non-Gaussian uncertainties. For example, distinguishing whether a spot is part of a galaxy or a foreground object adds further variance to the residuals. Thus, while our model is effective in estimating Gaussian-distributed residuals, extending it to handle these more intricate scenarios is an avenue for future work.

In machine learning, to estimate the quantity of interest following the setting above, we can employ a mean variance estimation (MVE) network \citep{Nix1994}. These models use a negative Gaussian loglikelihood as a loss function and predict an estimate of the quantity and their uncertainties. Despite its fame, several authors have observed that training an MVE network can be challenging (see, e.g. \citealp{ Detlefsen2019, Seitzer2022, sluijterman2023}). The primary argument is that the network tends to focus on areas where it performs well, consequently neglecting initially poorly fitted regions. 
One of the most common solutions for this problem is the application of an ensemble methodology; the main architecture using this procedure is the widely used deep ensembles regression \citep{Lakshminarayanan2017} that uses an ensemble of MVE to mitigate this effect. 

In the following section, we first present how the state-of-the-art methodology, deep ensembles, would function in predicting the features of interest and their uncertainties. Subsequently, we demonstrate how our two-step network better predicts the features of interest and addresses the controversial problem of the challenging training of MVE networks. In addition, we compare our results with the ones of SourceExtractor applied to the synthetic images.

\subsection{Deep ensemble regression on synthetic images}
\label{sec: Deep_ensemble}

Deep neural networks (DNNs) are prevalent black box predictors demonstrating robust performance across various tasks. However, determining the uncertainty of predictions made by DNNs remains a challenging problem yet to be fully resolved. Uncertainties are an inherent aspect of many real-world problems, and it is essential to understand the different types of uncertainty that can arise when building models. In machine learning, uncertainties are usually divided into epistemic and aleatoric uncertainty \citep{kendall2017, Kiureghian2009}. Epistemic uncertainty refers to the uncertainty that arises from a lack of knowledge, while aleatoric uncertainty refers to uncertainty that arises from stochastic processes and that cannot be reduced no matter how good our model is. Epistemic uncertainties can be mitigated and, if a model has enough information and is well-calibrated, can become negligible \citep{Smith2018, Wilson2020}.

One of the standard solutions to mitigate epistemic uncertainties is employing Bayesian DNNs, which estimate uncertainty by modelling a distribution over the network's weights \citep{kendall2017,blundell2015}; this approach necessitates significant modifications to the training process and is computationally expensive. Another often-used method to address this issue, deep ensemble regression, offers an alternative, user-friendly approach requiring no modification to the model or loss function: an ensemble of MVE networks. Training multiple MVE networks on identical data, predicting their $\hat{\beta}$ and $\hat{\Sigma}$, and then combining their predictions is a straightforward strategy for enhancing the MVE network's predictive uncertainty estimation and also partially mitigating the MVE training problems introduced in Section \ref{sec: Intro}. 

We trained three MVE networks, a small deep ensemble setting, on the problem at hand. Each network outputs six values: the predicted values for flux, x, and y, and their uncertainties. It is worth noting that while we do not estimate the full covariance matrix due to the primary quantities (flux, x, and y) being mostly uncorrelated within a cutout, our model can be easily extended to account for the full covariance matrix. This would involve increasing the output to nine values: three for the quantities and six for the covariance matrix components. 

The training process is straightforward and employs the negative loglikelihood as a loss \citep{Lakshminarayanan2017}:

\begin{equation}
\text{NLL} = \frac{1}{2}\log|\boldsymbol{\Sigma}| + \frac{1}{2}(\boldsymbol{\beta}-\boldsymbol{\hat{\beta}})^T\boldsymbol{\Sigma}^{-1}(\boldsymbol{\beta}-\boldsymbol{\hat{\beta}}).
\end{equation}

\begin{figure}
    \centering
    \includegraphics[scale=0.28]{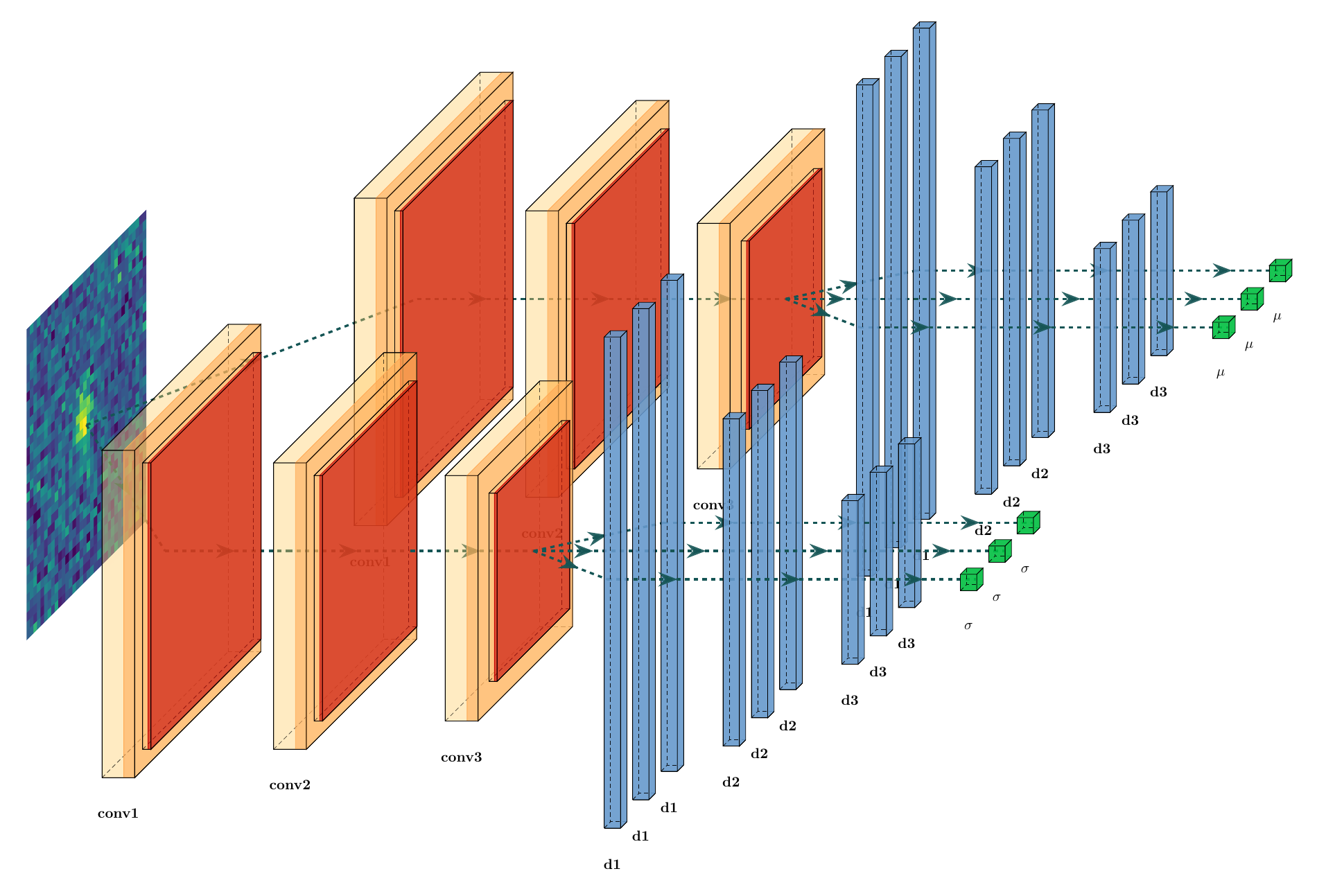}
    \caption{MVE architecture used for the deep ensemble. Two branches of three convolution layers followed by three independent groups of dense layers, one for each variable of interest. Inputs are single-band optical images; outputs are $\hat{\beta}$ and $\hat{\Sigma}$.}
    \label{fig: Architecture_ensemble}
    
\end{figure}

The architecture comprises two branches: one for the means and another for the uncertainties. Each branch contains three CNN blocks for feature extraction from the images and three dense layers independently trained for each feature. The CNN blocks in each branch consist of 4x4 kernel-sized convolutional layers with 32 channels, followed by max-pooling. No dropout layers are used in the CNN blocks. After the feature extraction, the multilayer perceptron (MLP) consists of three hidden layers with 128, 64, and 32 neurons, respectively.
We trained the networks using the Adam optimiser with a learning rate of 0.001 and a learning rate decay of 0.1. The training process used early stopping with patience of 8 epochs to prevent overfitting. As introduced in Sec. \ref{sec: Synthetic_dataset}, we also applied data augmentation techniques to improve the model's generalisation capabilities.
This design allows for more efficient learning of both the mean predictions and their associated uncertainties, contributing to the overall performance and accuracy of the proposed method. The entire architecture is shown in Fig. \ref{fig: Architecture_ensemble}.

\begin{figure}
    \centering
    \includegraphics[scale=0.44]{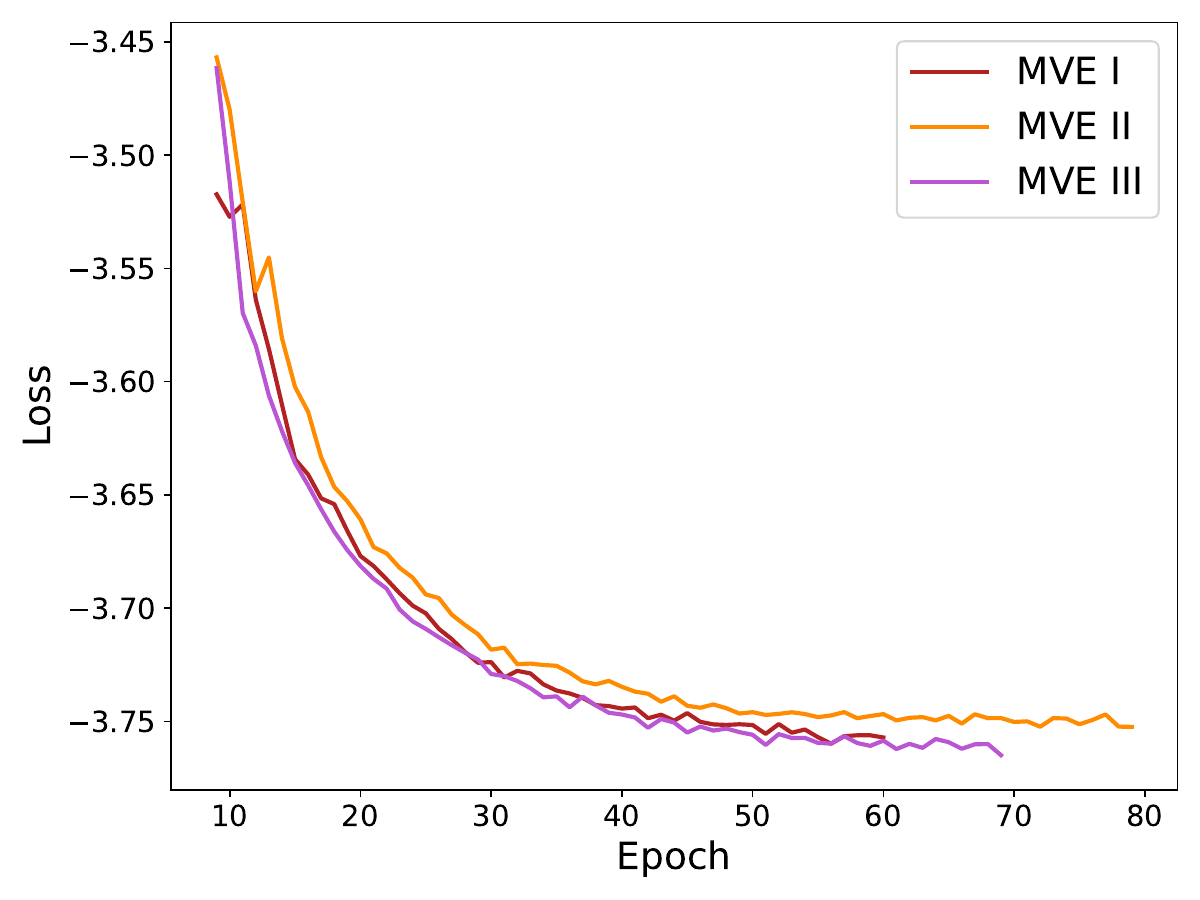}
    \caption{Training negative gaussian loglikelihood loss for the three models with different initialisations: MVE I, MVE II, and MVE III. The plot shows the loss values computed over multiple training iterations on the dataset. Results suggest that MVE III outperforms the other models in minimising the loss function. However, all three models are used being an ensemble setting.}
    \label{fig: Ensemble_loss}
\end{figure}

The deep ensemble MVE method is based on replicating the training process multiple times; this can provide a more accurate distribution of estimated parameters and uncertainties. However, as illustrated in Fig. \ref{fig: Ensemble_loss}, the outcomes vary and converge to distinct loss values depending on the initialisation of the model's weights. 

\noindent
This sensitivity of MVE networks to the random initialisation values of the weights can lead to suboptimal solutions. An ensemble of the MVE networks, averaging the predictions from the three MVEs, results in a more robust estimation of the target being less dependent on a single random initialisation. Nevertheless, it is essential to note that the deep ensemble methods require increased computational resources due to the need to train multiple models. 

\noindent
In this paper, we propose to solve the problem at its core, improving the trainability of MVEs, and ensuring better performances and more consistent results.

\subsection{Two-step network on synthetic images}
\label{sec: Two_steps_network}

In this section, we introduce the two-step mean variance estimation (TS-MVE) approach, an expansion of the method presented in \cite{sluijterman2023} for more effective training of MVE networks in regression tasks. This novel TS-MVE approach has been specially adapted for image feature regression, confronting the complex challenge of estimating uncertainties while simultaneously delivering enhanced accuracy. 

The solution for an improved training approach such as TS-MVE stems from recent insights shared by \cite{sluijterman2023}. Their study underscores a crucial flaw in MVE networks when the parameters $\beta$ and $\Sigma$ are trained concurrently. The highlighted challenges encompass instability in training MVE networks and a tendency for the network to falter in learning the mean function, particularly in regions where it initially has a large error. This predicament can trigger an escalation in the variance estimate, driving the network to disproportionately concentrate on well-performing regions and neglect areas of poor fit.

Our novel TS-MVE method strives to overcome these identified limitations. Our key objectives with this two-step network are twofold. Firstly, we aim to achieve lower mean squared error (MSE), mean absolute error (MAE), and negative loglikelihood (NLL) values than those yielded by conventional MVE networks. Secondly, we intend to demonstrate that, regardless of the initial parameter settings, our TS-MVE model exhibits consistent stability, converging reliably towards a more optimised loss value.

The proposed TS-MVE method maintains the same number of parameters as a single iteration of the MVE architecture shown in Fig. \ref{fig: Architecture_ensemble}, ensuring the model's complexity remains comparable. However, the key differentiating factor of our approach is the implementation of two distinct training stages, which leads to a more robust and effective training process.
In the first stage, the model focuses on learning the mean function of the target variable. This stage primarily emphasises capturing the overall structure and trends present in the data, establishing a solid foundation for subsequent variance estimation. The primary objective of this stage is to minimise the Huber loss \citep{Huber1964,hastie2009}, a widely used and reliable metric for regression tasks as shown in Eq. \ref{eq: Huber}. 
Following the completion of the first stage, the second stage begins, which involves learning the variance function of the target variable. By incorporating the information gathered during the initial stage, the model can now concentrate on estimating the uncertainties of its predictions. This is achieved through a negative Gaussian loglikelihood loss function, which allows the model to effectively capture and represent the complex relationships between features and uncertainties in the target variable.

\noindent
The separation of mean and variance estimation allows for a more focused learning process, enabling the model to accurately capture the underlying structure of the data while accounting for uncertainties.

\subsubsection{Part I: estimating $\beta$}

\begin{figure}
    \centering
    \includegraphics[scale=0.3]{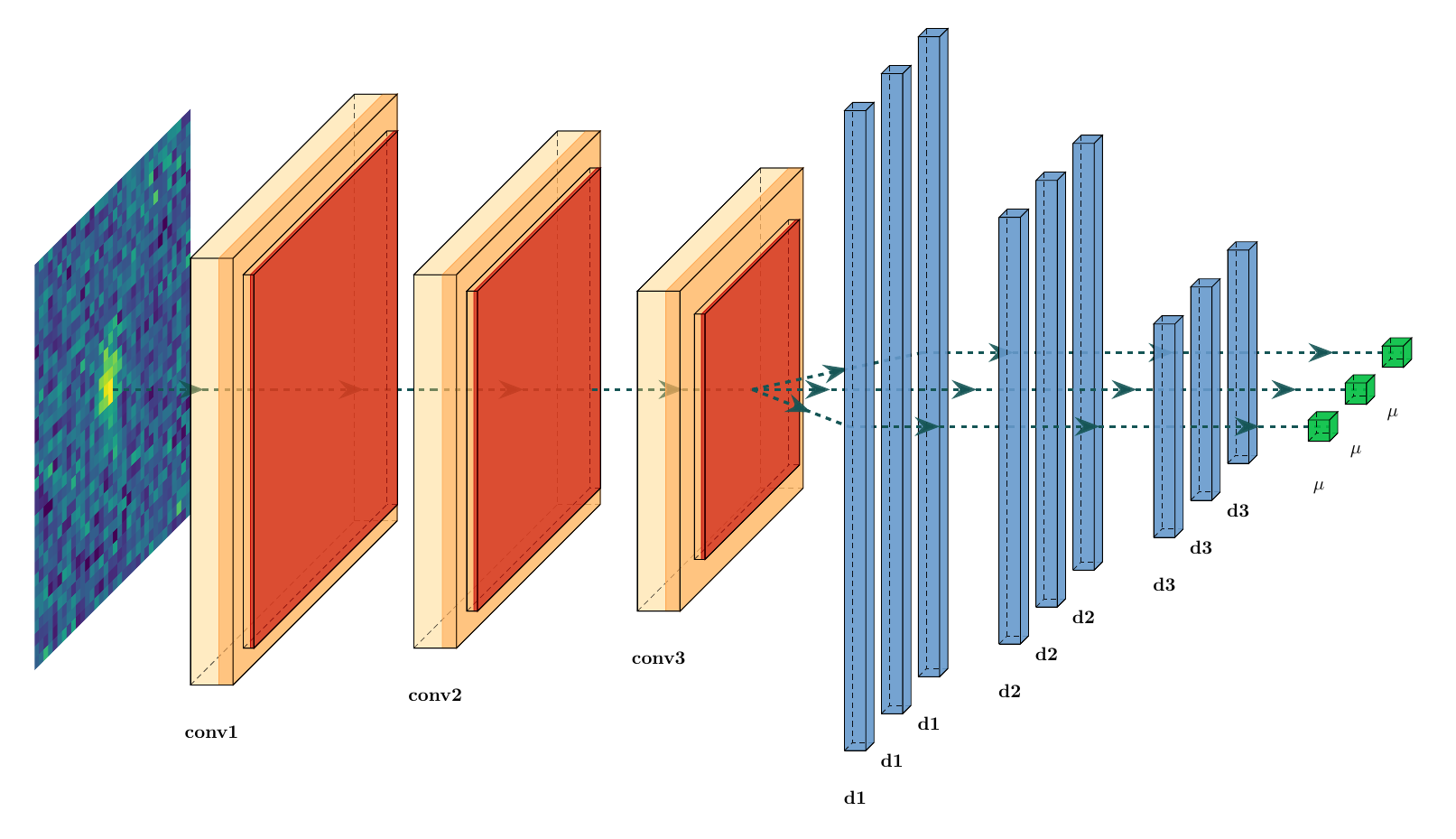}
    \caption{Architecture used for Part I network. Same as a single branch of Fig. \ref{fig: Architecture_ensemble}: three convolution layers followed by three independent groups of dense layers, one for each variable of interest. Inputs are single-band optical images; output is $\mu$ of flux, x and y. }
    \label{fig: Architecture_PartI}
    
\end{figure}

In the first stage of the TS-MVE approach, we train a network that estimates the target variables' mean values, $\beta$. This network is primarily concerned with learning the underlying relationships between the input features and the target variables, x, y, and flux, without considering uncertainties. The architecture of the Part I network, as depicted in Fig. \ref{fig: Architecture_PartI}, follows the same structure as a single branch of the MVE, shown in Fig. \ref{fig: Architecture_ensemble}.

The Part I network is trained to minimise the Huber loss :

\begin{equation}
\label{eq: Huber}
  L_\delta(y,f(x))=\begin{cases} 
\frac{1}{2}(y-f(x))^2 & \text{for} \; |y-f(x)| \leq \delta \\
\delta(|y-f(x)|-\frac{1}{2}\delta) & \text{otherwise}
\end{cases}  
\end{equation}


\noindent
The Huber loss represents a hybrid of the MSE and MAE loss functions, rendering it less sensitive to outliers than a simple MSE loss function. By adjusting the value of the parameter $\delta$, the user can modify the balance between the two types of loss functions, optimising performance according to the specific demands of the dataset. For our application, we use the standard value of $\delta = 1$. 
While the Huber loss is not directly comparable to the loss of the ensemble network, we observe that the MSE and MAE values for the TS-MVE Part I network are lower than those of any of the MVEs of the ensemble. A comprehensive comparison of the performance metrics of each network is provided in Section \ref{sec: Metrics_comparison}.

\subsubsection{Part II: estimating $\Sigma$}

\begin{figure}
    \centering
    \includegraphics[scale=0.28]{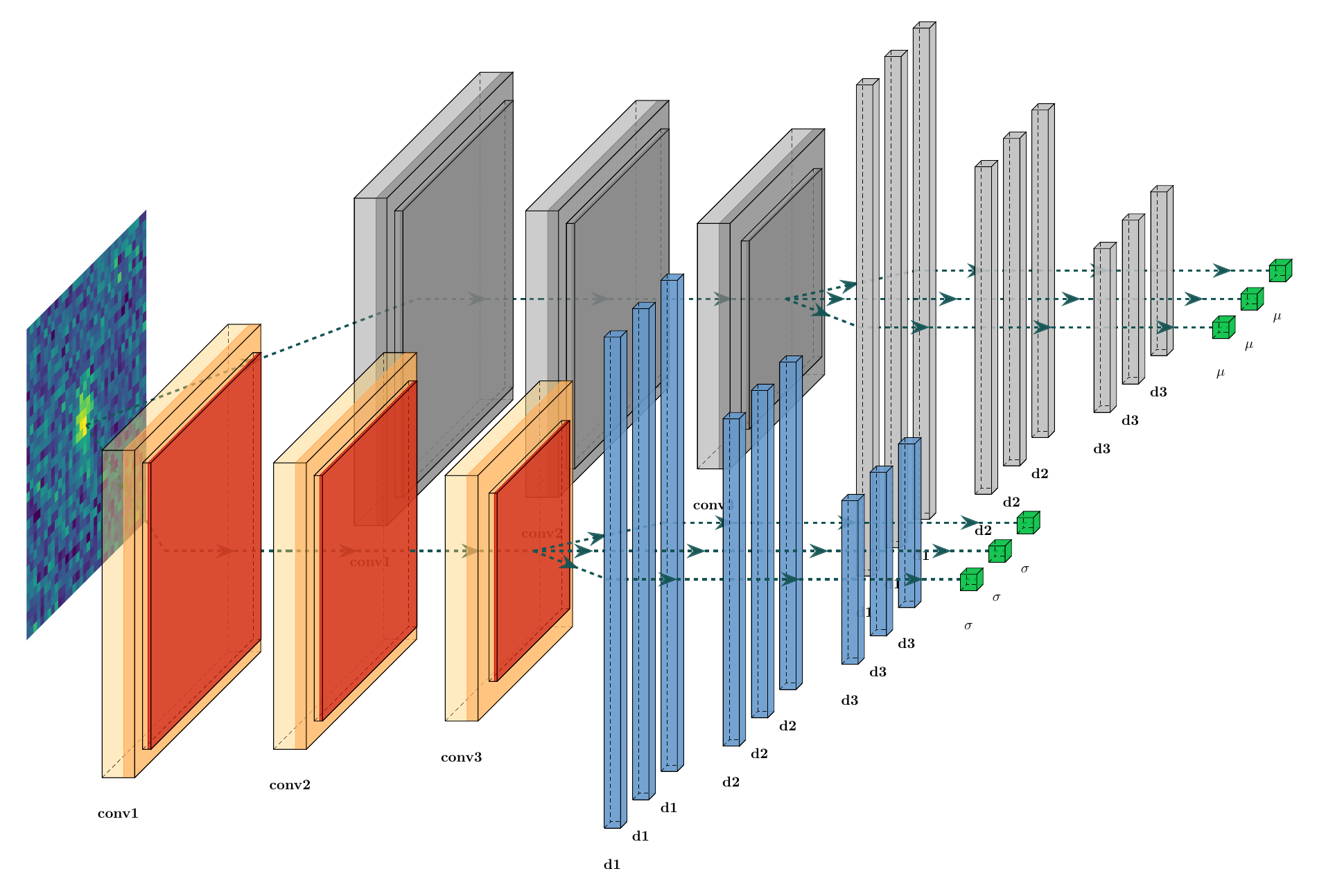}
    \caption{Architecture used for Part II network. Same as shown in Fig. \ref{fig: Architecture_ensemble}: two branches of three convolution layers followed by three independent groups of dense layers, one for each variable of interest. The grey branch is not trained from scratch but is initialised with the results of the Part I network. Inputs are single band optical images; outputs are improved $\mu$ and predicted $\sigma$ of flux, x and y.}
    \label{fig: Architecture_PartII}
    
\end{figure}

In the second stage of the TS-MVE approach, we leverage the optimised parameters obtained from the Part I network to predict the uncertainties, $\Sigma$, using a second network. The architecture of the Part II network, as illustrated in Fig. \ref{fig: Architecture_PartII}, uses the same structure of the complete MVE architecture presented in Fig. \ref{fig: Architecture_ensemble}. The key distinction lies in that the weights for the mean branch are already set to the optimised values derived from the Part I network. The objective of the Part II network is to predict the most suitable $\sigma$ values while concurrently enhancing the mean prediction. In the present study, we assume no correlation exists between the variables x, y, and flux. Nevertheless, the complete covariance matrix $\Sigma$ prediction is feasible and straightforward, presenting no significant complications.

During the Part II network training, we employ the negative Gaussian loglikelihood loss as the loss function. This loss function is particularly suited for uncertainty estimation, as it accounts for both the difference between the predicted mean and the true target value and the predicted uncertainty.

\noindent
By dividing the training process into two stages, TS-MVE can effectively learn the target parameters and their associated uncertainties more stably. This separation of concerns allows each network to focus on a specific aspect of the problem, resulting in improved overall performance.

\subsubsection{Metrics comparison}
\label{sec: Metrics_comparison}

In this section, we evaluate and compare the performance of the TS-MVE network with the MVEs of the ensemble using two widely adopted metrics, mean squared error (MSE) and mean absolute error (MAE). These metrics offer a quantitative assessment of the accuracy and precision of the models.

\begin{figure}
    \centering
    \includegraphics[scale=0.44]{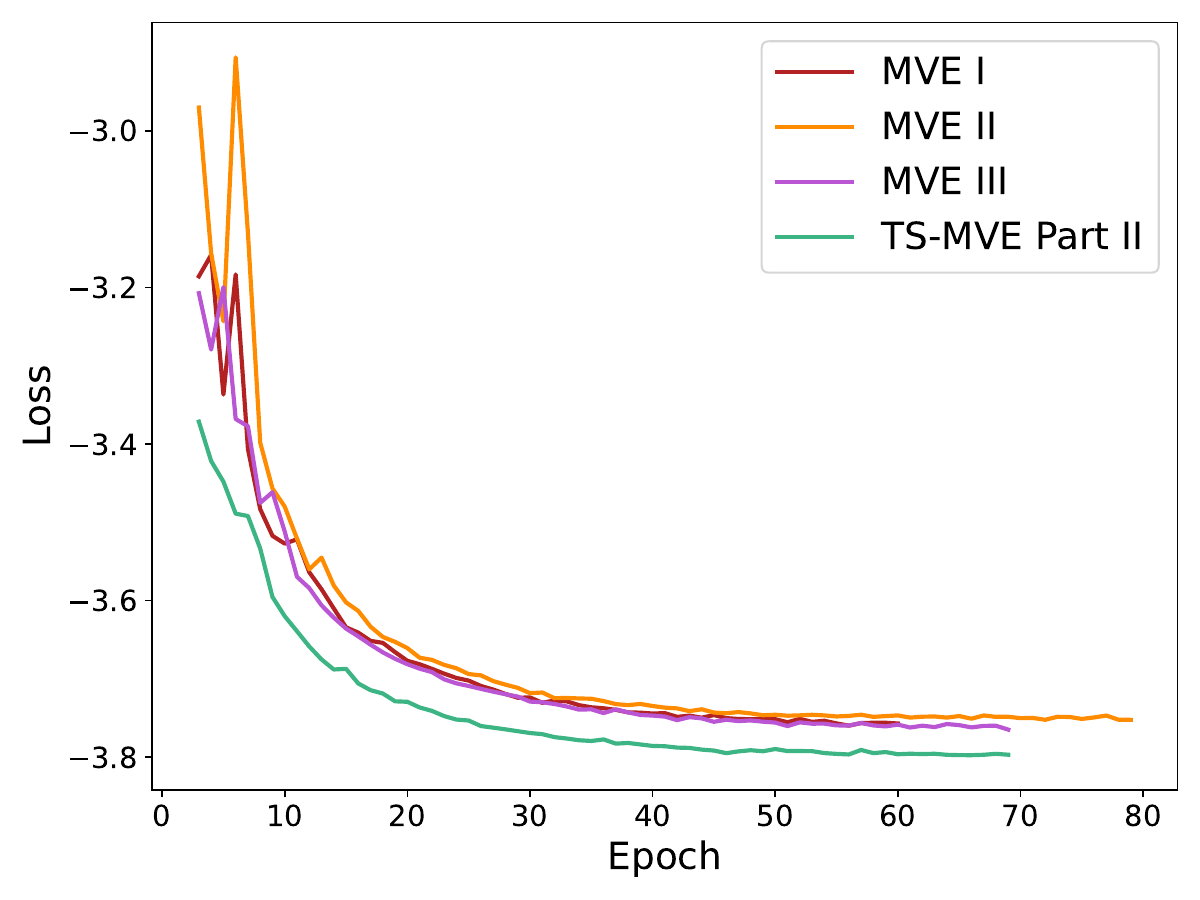}
    \caption{Comparison of the loss for the TS-MVE Part II and all the MVEs models part of the ensemble. The TS-MVE converges to a lower loss value than any of the MVEs.}
    \label{fig: Comparison_loss}
\end{figure}

\begin{figure}
    \centering
    \includegraphics[scale=0.44]{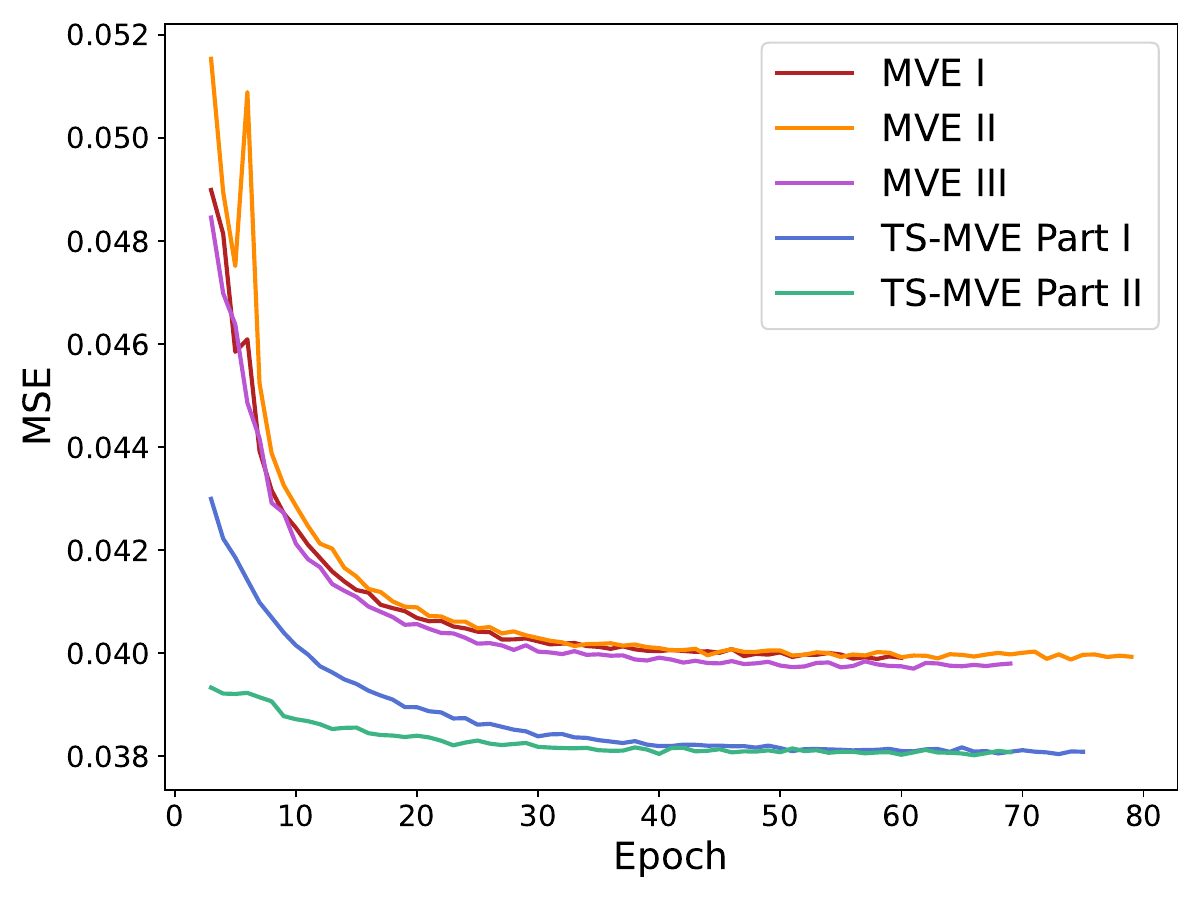}
    \caption{Comparison of the MSE metric for all models. TS-MVE Part I and Part II network predictions for $\mu$ have significantly lower MSE than any of the MVEs composing the ensemble network.}
    \label{fig: Comparison_MSE}
\end{figure}

\begin{figure}
    \centering
    \includegraphics[scale=0.44]{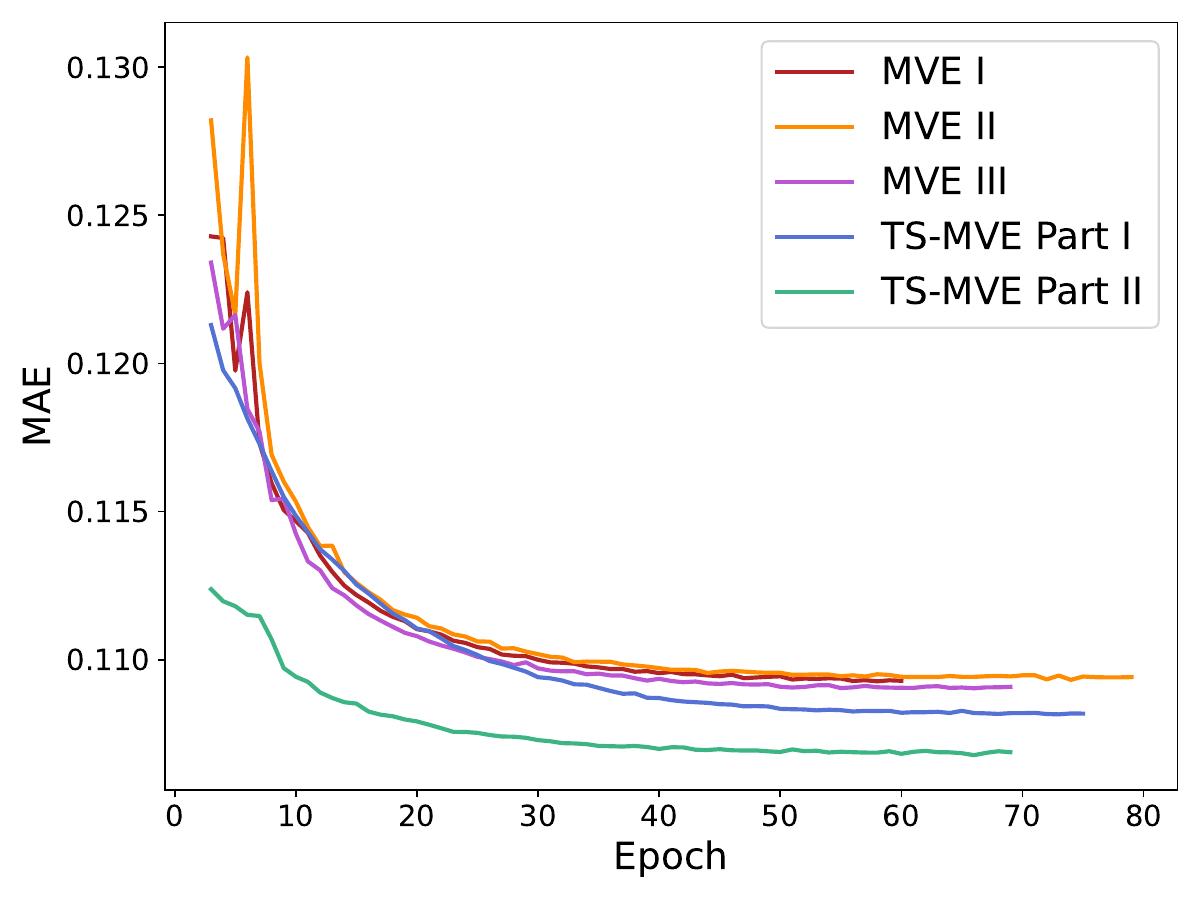}
    \caption{Comparison of the MAE metric for all models. TS-MVE Part I and Part II network predictions for $\mu$ have significantly lower MAE than any of the MVEs composing the ensemble network. TS-MVE Part II, built on the results of Part I, converges to an even lower MAE value.}
    \label{fig: Comparison_MAE}
\end{figure}

Our analysis highlights several advantages of the TS-MVE approach over the ensemble models. As shown in Fig. \ref{fig: Comparison_loss}, \ref{fig: Comparison_MSE}, and \ref{fig: Comparison_MAE}, the TS-MVE network achieves lower MSE, MAE, and NLL values, indicating superior overall accuracy. This improvement in both MSE and MAE metrics demonstrates the effectiveness of the TS-MVE method in reducing prediction errors and enhancing precision. 
Our analysis also revealed that the TS-MVE network exhibits less dependence towards random weight initialisation, demonstrating no significant differences between models with varying initialisations. To illustrate, when training the Part I network multiple times with different initialisations while maintaining the same architecture, we noticed a uniformity in the models converging towards a consistently small Huber loss value. Similarly, the corresponding Part II networks trained under different initial conditions also converged towards nearly identical negative loglikelihood values. This remarkable consistency underscores the robustness of the TS-MVE network against variability in initial weight settings.

Our findings suggest that, for feature regression from images, the TS-MVE network indeed outperforms a standard MVE network. Interestingly, an ensemble composed of the three MVE networks also did not offer any performance improvement, despite being more computationally expensive. However, this does not signify an overarching redundancy of deep ensemble techniques. Quite to the contrary, in certain situations, these techniques remain crucial. Notably, even when the model demonstrates stability, some epistemic uncertainty may persist, such as when comprehensive knowledge about the underlying system is lacking \citep{Rasmussen2004}.
Nevertheless, it is evident that a TS-MVE network tends to yield superior performance and offers a stronger buffer against epistemic uncertainties compared to its conventional counterparts. Furthermore, it is important to note that the TS-MVE approach is not mutually exclusive with ensemble techniques. Despite the strength of a single TS-MVE model, we can further enhance performance and robustness by employing an ensemble of TS-MVE networks. This combination could potentially offer a powerful tool for addressing complex regression tasks, combining the specific advantages of the TS-MVE approach with the general benefits of ensemble methods.

\subsubsection{Saliency maps}
\label{sec: Saliency_maps}

Prediction models in machine learning can be highly reliable when developed and deployed appropriately. However, the reliability of a model depends on various factors, such as the quality and quantity of data used to train the model, the choice of algorithms and parameters, and the complexity of the problem. Thus, it is essential to evaluate the reliability of a prediction model before using it in real-world applications.

One way to assess the reliability of a machine learning model is by studying its output. The output of a model can be examined on the performance of the test dataset as in the previous sections, but it can also be examined based on the relationship between input and output.
Deep taylor decomposition (DTD) and axiomatic attribution are methods used to explain decisions made by nonlinear classification models such as deep neural networks. DTD decomposes the decision-making process into simpler components, enabling a better interpretation of the model's predictions \citep{Montavon2017}. Axiomatic attribution provides a mathematical framework to understand the contribution of individual input features to the final prediction \citep{Sundararajan2017}.

For image classification and regression, some specific methods have been developed to evaluate the reliability of networks, Saliency Maps \citep{Simonyan2013}. They are commonly used to visualise and understand the behaviour of machine learning models, particularly CNNs. A saliency map visualises which parts of the input image are most important for a given prediction by the model. There are several different methods for creating saliency maps. One common approach is applying a gradient-based optimisation method to the input to find the pixels that significantly impact the model's output. 

In particular, we used SMOOTHGRAD \citep{Smilkov2017}, a method for creating saliency maps that aim to reduce noise and improve the interpretability of the resulting maps. Saliency maps can sometimes be noisy, with little and unimportant features highlighted as important due to random fluctuations in the model's output. SMOOTHGRAD addresses this problem by creating an ensemble of saliency maps, each slightly perturbed with Gaussian noise from the original input. The resulting maps are then averaged together, which helps to smooth out the noise and highlight more robust and reliable features.

\begin{figure}
\centering
\includegraphics[scale=0.2]{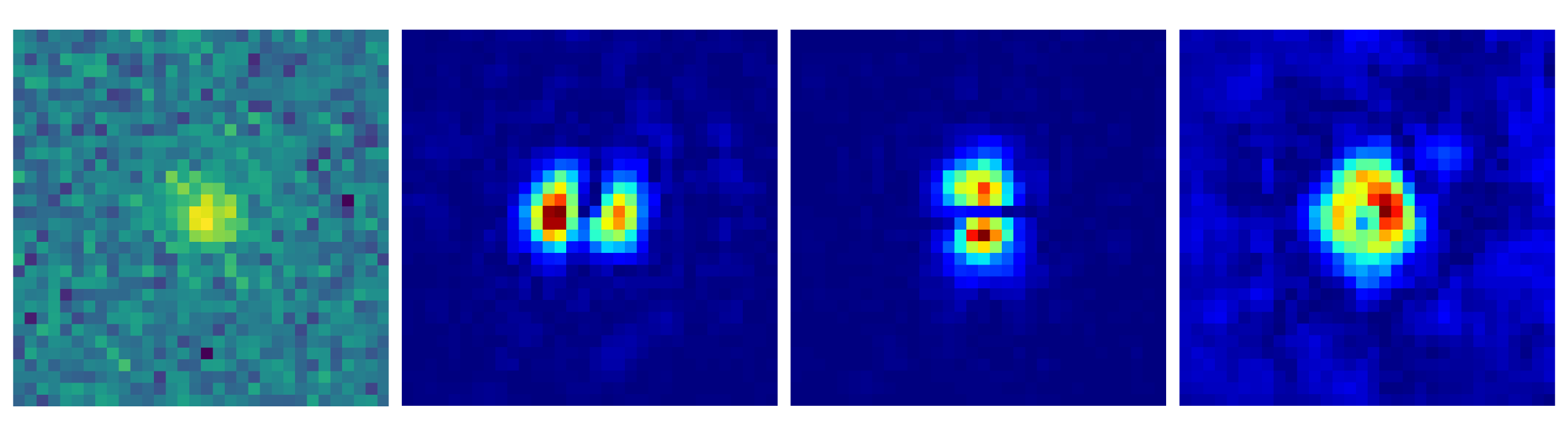}
\caption{Saliency map of ASID-FE's features prediction for a single source. The images are from left to right: original image, saliency for $x$, saliency for $y$, and saliency for $flux$. The colour map indicates the degree of importance assigned to each pixel in the image. The brighter the colour, the more important the pixel is for the prediction. As expected, the network mainly focuses on the source to determine its centre and flux while paying less attention to the edges of the image.}
\label{fig: Saliency_map}
\end{figure}

In Fig. \ref{fig: Saliency_map}, we show the saliency map for predicting a single source of ASID-FE. The network seems to be able to discern the source from the background, focusing primarily on the source to predict its x, y, and flux properties. Interestingly, the network pays less attention to the edges of the image, which suggests that it has learned to differentiate between relevant and less relevant information for the task at hand.

\noindent
This observation highlights the network's capability to recognise and prioritise the relevant features of the input image for accurate prediction, thereby demonstrating its effectiveness in handling the task. The dipole-like structure observed in the saliency maps for $x$ and $y$ could be attributed to the network learning the gradient direction in the source's position, which is an important factor in determining the source's centre. Further investigation is needed to understand better this structure's significance and implications on the network's predictions.

\begin{figure}
\centering
\includegraphics[scale=0.2]{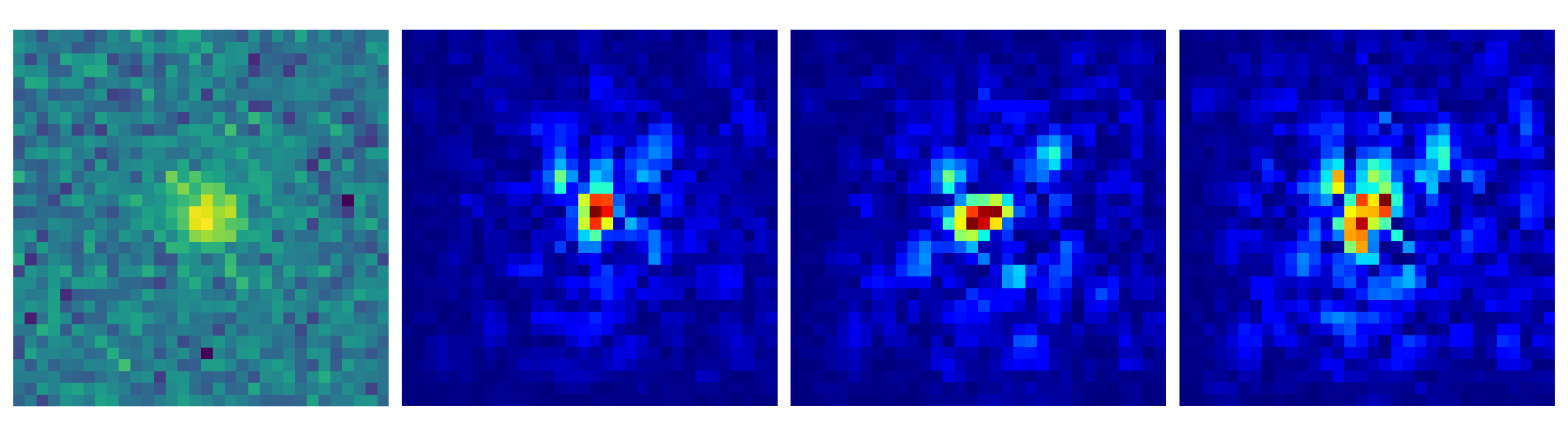}
\caption{Saliency map of ASID-FE's uncertainties prediction for a single source. The images are from left to right: original image, saliency for $\sigma_x$, saliency for $\sigma_y$, and saliency for $\sigma_{flux}$. The colour map indicates the degree of importance assigned to each pixel in the image. The brighter the colour, the more important the pixel is for the prediction. Results suggest that the network focuses mostly on the central pixels to estimate $\sigma_x$ and $\sigma_y$, while it pays more attention to a wide area around the source to estimate $\sigma_{flux}$.}
\label{fig: Saliency_map_uncert}
\end{figure}

\noindent
In Fig. \ref{fig: Saliency_map_uncert}, we present the saliency map for the uncertainties of the three quantities of interest: $\sigma_x$, $\sigma_y$, and $\sigma_{flux}$. The visualisation shows that the network adopts a different approach when estimating uncertainties compared to predicting the main features. Notably, the network appears to place more importance on the background pixels than before, which suggests that the background may contain valuable information for estimating uncertainties.
For $\sigma_x$ and $\sigma_y$, the network seems to focus primarily on the central pixels of the source. This could imply that the model is leveraging the central region's intensity distribution to estimate the positional uncertainties. In the case of $\sigma_{flux}$, the network pays more attention to a wider area around the source, which may indicate that the network highly considers the source's background when estimating the uncertainty in flux.

These observations provide valuable insights into the neural network's decision-making process when estimating uncertainties. By understanding how the network focuses on different regions and features of the input image, we can identify potential areas of improvement and develop strategies to enhance the network's accuracy and robustness.

\section{Results}

This section offers a comprehensive evaluation of ASID-FE's performance in predicting the flux and location of sources within astronomical images, alongside a comparison with SourceExtractor \citep{Bertin1996}. As a widely used software package in the field of astronomy, SourceExtractor has become a standard tool for detecting and analysing celestial objects such as stars and galaxies. Our analysis is divided into two main subsections: the first focuses on a detailed examination of ASID-FE and SourceExtractor using synthetic test sets (comprising approximately 600k sources), and the second extends the evaluation to real-world data, specifically MeerLICHT images.

The synthetic test set analysis explores the accuracy, uncertainties, and behaviour of the two methods across different signal-to-noise levels, employing both quantitative metrics and visualisations. The real-world data analysis assesses the models' generalisation, adaptability, and alignment with established software. Together, these evaluations provide a robust understanding of ASID-FE's strengths and limitations and its comparative performance with SourceExtractor in the context of astronomical image analysis.

\subsection{Results of ASID-FE and SourceExtractor on synthetic images}
\label{sec: SourceExtractor_synthetic}

In this subsection, we assess the performance of ASID-FE and SourceExtractor in predicting the flux and location of sources within our synthetic test set of astronomical images. Each source in the test set is accompanied by the predicted flux and centre coordinate as determined by SourceExtractor, applied to full-field simulated images.

Our analysis begins with a comparison of the methods' predictions for the x and y coordinates. We only present results for the x coordinate as the y coordinate showed no significant difference in performance. SourceExtractor's inherent flexibility offers a variety of estimators for parameter determination, and we selected the windowed positions to estimate source locations. 
For the estimation process, SourceExtractor determines the source position through an iterative Gaussian fitting approach that refines the source profile until a definitive location is identified. Following this, uncertainties in these positions are calculated using a windowed centroiding method. Based on the assumption of uncorrelated pixel noise, standard error propagation is applied to derive variances and covariance for the windowed $x$ and $y$ coordinates. These are obtained by taking weighted sums of the squared deviations between each pixel's position and the overall windowed centroid, and then normalising by the square of the weighted intensity sum. This approach yields a robust quantification of positional uncertainties.

\begin{figure}[h]
        \includegraphics[width=0.48\textwidth]{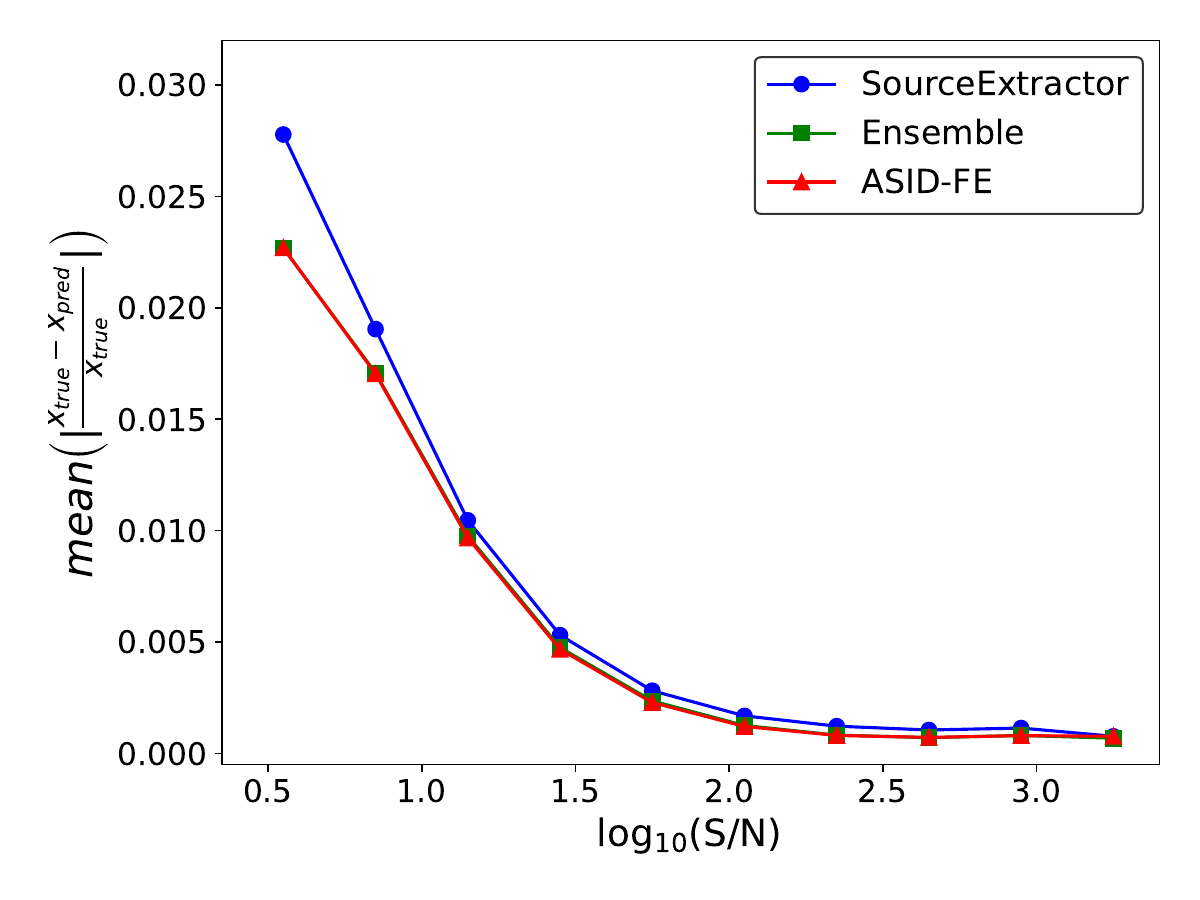}
\caption{Mean absolute relative error for the $x$ coordinate by SourceExtractor (blue), the MVEs ensemble (green) and ASID-FE (red). ASID-FE and the ensemble have overall better results than SourceExtractor at any S/N level.}
\label{fig: X_MARE}
\end{figure}

The results of this comparison are presented in Fig. \ref{fig: X_MARE}, where we examine the mean absolute relative error (MARE) with respect to the true synthetic values as a function of S/N:

\begin{equation}    
\centering
MARE =  \frac{1}{n} \sum  \left|\frac{\beta - \hat{\beta}}{\beta}\right| .
\label{eq: MARE} 
\end{equation}
\noindent
Here, $\beta$ represents the true value, $\hat{\beta}$ represents the estimated value by the algorithms, and the sum is over the sources within an S/N bin.

Our evaluations reveal an astonishing level of accuracy in predicting sources' x and y positions across all three methods. While ASID-FE and the ensemble slightly outperform SourceExtractor, the differences are subtle. What sets ASID-FE apart is its efficiency: achieving comparable results as a single network, it stands as a more computationally streamlined solution compared to the ensemble, which is composed of three MVE networks of the same size. This efficiency does not compromise performance, making ASID-FE a compelling choice in this context.

\begin{figure}[h!]
     \includegraphics[width=0.48\textwidth]{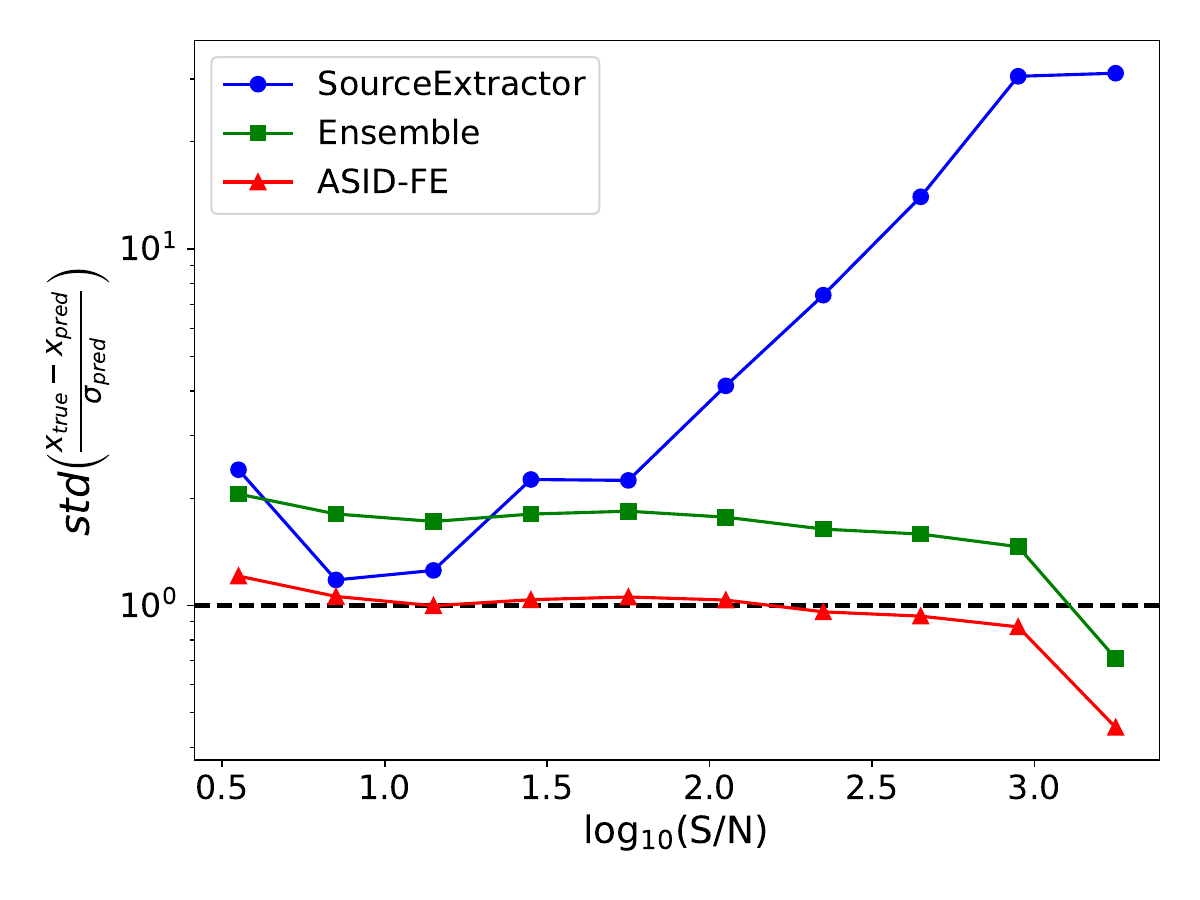}
     \caption{Standard deviation of the standardised residuals for the x coordinate by SourceExtractor (blue), the MVEs ensemble (green) and ASID-FE (red). ASID-FE has overall better results than SourceExtractor at any S/N level.}
\label{fig: X_WRSD}
\end{figure}

Next, we delve into the uncertainties associated with these predictions. We utilise the standard deviation of the standardised residuals as a robust metric for evaluating the performance of both estimator and uncertainties. Ideally, this metric should be close to 1, indicating that the residuals are approximately normally distributed and that the estimator's uncertainty is well-calibrated. Deviations from 1 suggest that the estimator is either overconfident or underconfident in its predictions. In Fig. \ref{fig: X_WRSD}, we present the standard deviation of the standardised residuals for the $x$ coordinate as estimated by SourceExtractor, the MVEs ensemble, and ASID-FE across various S/N levels. 

Some nuances in the calibration of uncertainties are observed across the different methods. SourceExtractor shows decently calibrated uncertainties at low S/N levels but tends to increasingly underestimate the uncertainties as the S/N rises. In particular, SourceExtractor's sigma values at high S/N can reach extremely small values, equivalent to approximately $0.0011$ arcseconds. These values are significantly smaller than the telescope's resolution of $0.56$ arcseconds per pixel, highlighting its tendency to underestimate uncertainties. 
The ensemble method performs better across the S/N spectrum but still tends to slightly underestimate $\sigma$. In contrast, ASID-FE exhibits excellent calibration, almost perfectly aligning at 1 for the standard deviation of the standardised residuals across most S/N levels. The only exception is an overestimation of uncertainties in the highest S/N bin, which is likely attributable to the scarcity of training samples at these levels. However, it is worth noting that these discrepancies are not highly consequential for any of the three algorithms, given the already high accuracy of the predictions themselves.

Having assessed the accuracy and uncertainties in predicting the spatial coordinates of the sources, we now shift our focus to another vital aspect of astronomical image analysis: the estimation of flux. Flux measurement is central to understanding the intensity and distribution of light from astronomical objects, and it requires a different set of considerations and methodologies compared to spatial localisation.
For this purpose, we compare our results with the FLUX\_AUTO output of SourceExtractor. This method estimates the flux by integrating pixel values within an adaptively scaled aperture, following Kron’s first-moment algorithm. Initially, it calculates the flux within an elliptical aperture, automatically defined to encompass most of the light from the source. The aperture's size and shape are dynamically adjusted based on the source's properties, such as brightness and spatial extent. This adaptability makes FLUX\_AUTO a robust and flexible method for measuring fluxes across various astronomical objects, expected to capture at least 90\% of the source flux.

\begin{figure}[h!]
     \includegraphics[width=0.48\textwidth]{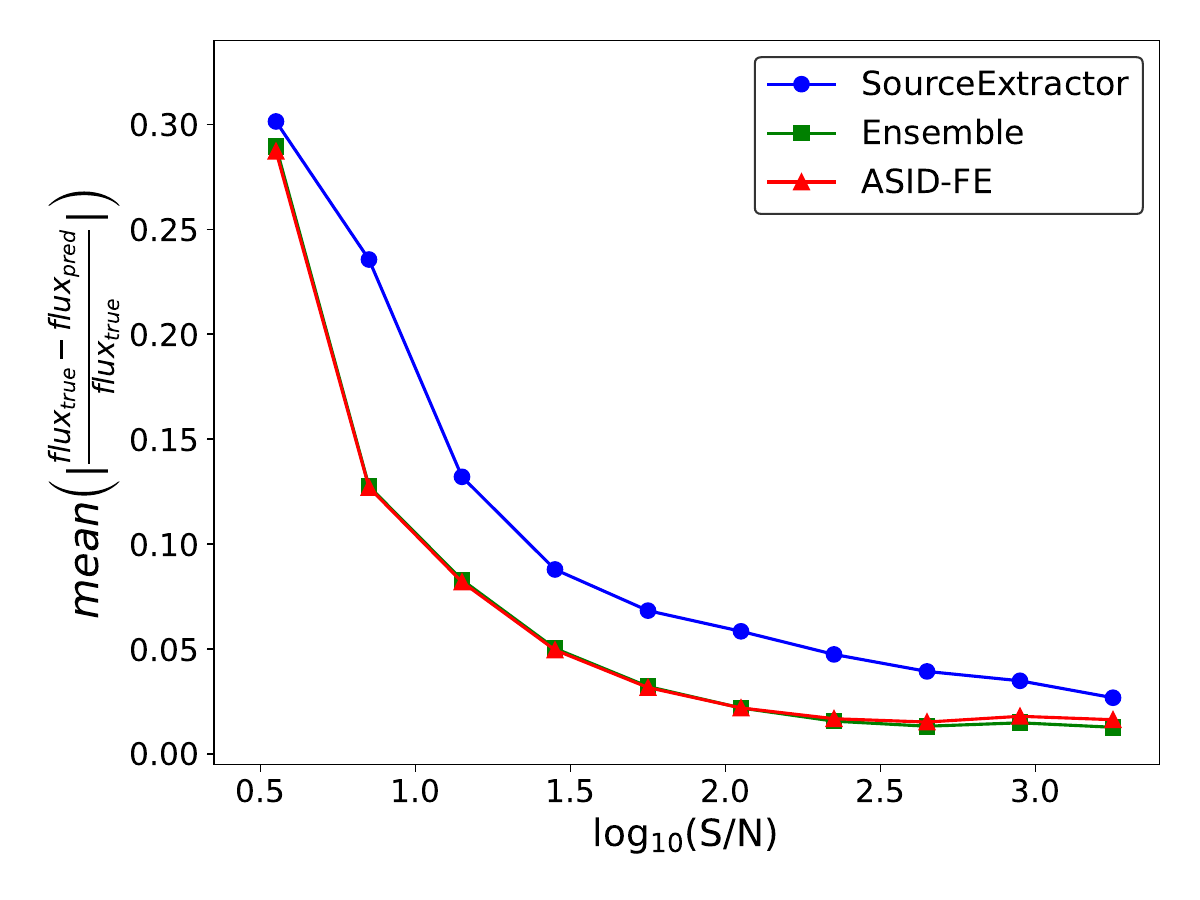}
     \caption{Mean absolute relative error for the flux by SourceExtractor (blue), the MVEs ensemble (green) and ASID-FE (red). ASID-FE and the MVEs ensemble have overall better results than SourceExtractor at any S/N level.}
\label{fig: Flux_MARE}
\end{figure}

The flux produced by SourceExtractor is directly comparable to the flux estimate from ASID-FE. In Figure \ref{fig: Flux_MARE}, we rigorously assess the flux estimates derived from SourceExtractor, ASID-FE, and the ensemble method, employing the mean absolute relative error as our evaluation metric across a spectrum of S/N.

\noindent
All three methods exhibit a decreasing MARE as the S/N increases. However, ASID-FE and the ensemble method consistently outperform SourceExtractor across the entire S/N range. The most pronounced difference is observed at medium S/N levels, where ASID-FE and the ensemble method demonstrate approximately 5-10\% less error compared to SourceExtractor. It is worth noting that the performance of ASID-FE at the extremes of the S/N spectrum is likely influenced by the scarcity of training samples at these levels, which could be an avenue for future improvement.

While the comparison of flux estimation yields promising results, a key difference arises in how uncertainty is quantified between SourceExtractor's FLUXERR\_AUTO and ASID-FE's uncertainty measure. This divergence is rooted not just in the mathematical equations used but also in the fundamental statistical interpretation of what uncertainty actually signifies. Although commonly referred to as an error in astronomical jargon, SourceExtractor's FLUXERR\_AUTO, from a statistical standpoint, is an estimate of the standard deviation of the true, unobservable flux \( F \) rather than an error in the flux estimator \( \hat{F} \). \\
To further elucidate this distinction, we shall consider the statistical problem at hand, which consists of three main components:
$$F \rightarrow F^* \rightarrow \hat{F}$$
In this framework, \( F \) is the true, constant flux of the source. The telescope measures \( F^* \), which is a realisation of \( F \) subject to Poisson noise due to the photon-counting process, as well as additional complexities introduced by the instrument, for example the PSF. SourceExtractor, the ensemble, and ASID-FE aim to estimate \( F^* \) through an estimator \( \hat{F} \). However, SourceExtractor overlooks the inaccuracies in the estimated \( \hat{F} \) and uses it as a perfect proxy of \( F^* \).
SourceExtractor then proceeds to calculate \texttt{FLUXERR\_AUTO}, an estimate of the standard deviation of the true, unobservable flux \( F \), as:
$$\hat{\sigma}(F) = \sqrt{\left(\sqrt{\hat{F}} \; \right) ^2 + \hat{\sigma}_{\text{bkg}}^2 }$$
where $\hat{\sigma}_{\text{bkg}}$ and $\sqrt{\hat{F}}$ are an estimate of the standard deviation of flux in the background and of the source, respectively. 
In contrast, ASID-FE adopts a more nuanced methodology, explicitly accounting for the estimator imperfections. This approach yields what ASID-FE refers to as its sigma values, which serve as a measure of the divergence between \( \hat{F} \) and \( F^* \). By incorporating this sigma term into the equation, ASID-FE provides a more comprehensive and realistic estimate for the standard deviation of the true flux \( F \). This estimate not only captures the inherent uncertainties in \( F^* \) but also includes the uncertainties associated with the estimator \( \hat{F} \) itself:
$$
\hat{\sigma}(F)  = \sqrt{\left(\sqrt{\hat{F}} \; \right) ^2 + \hat{\sigma}_{\hat{F}}^2}
$$
Here, $\sqrt{\hat{F}}$ represents the standard deviation of the measured source, and \( \hat{\sigma}_{\hat{F}} \) represents the uncertainty on the estimator \( \hat{F} \). Importantly, it should be noted that the uncertainty term \( \hat{\sigma}_{\hat{F}} \) in ASID-FE's formulation inherently includes the background noise, \( \hat{\sigma}_{\text{bkg}} \). By including an additional component, ASID-FE offers a more detailed and accurate reflection of the true uncertainties. The lack of this additional component in SourceExtractor's formulation might explain why its uncertainties are often underestimated for high S/N sources, as noted in studies such as \citeauthor{Becker2007} (\citeyear{Becker2007}) and \citeauthor{Sonnett2013} (\citeyear{Sonnett2013}).

\begin{figure}[h!]
     \includegraphics[width=0.48\textwidth]{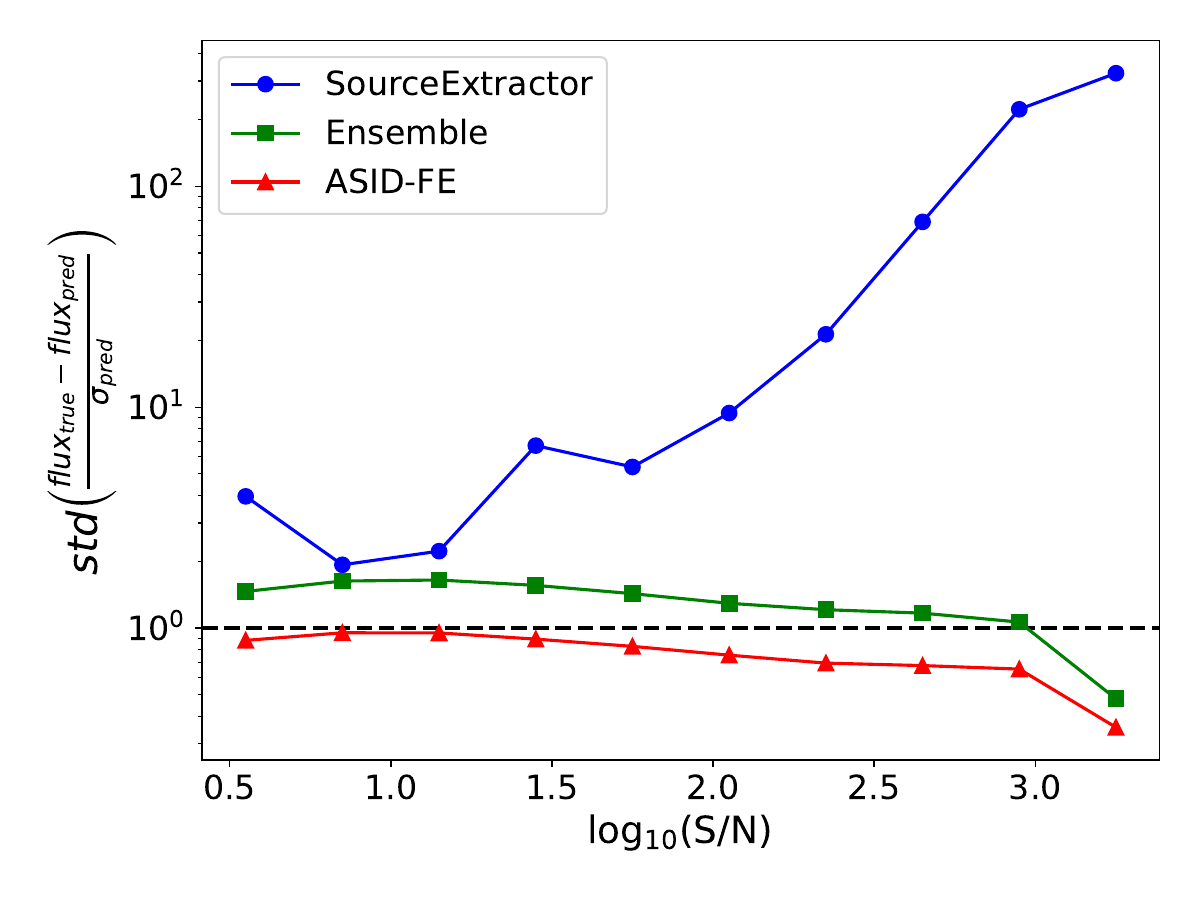}
     \caption{Standard deviation of the standardised residuals for the flux and $\hat{\sigma}(F)$ by SourceExtractor (blue), the MVEs ensemble (green) and ASID-FE (red). ASID-FE's results more closely adhere to the ideal value of 1, showing overall better-calibrated uncertainties at any S/N level.}
\label{fig: Flux_WRSD}
\end{figure}

In Fig. \ref{fig: Flux_WRSD}, we present the standard deviation of the standardised residuals for the flux as estimated by SourceExtractor, the MVEs ensemble, and ASID-FE across various S/N levels. 

\noindent
The plot reveals nuanced differences in the performance of the three methods. Specifically, at high S/N levels, SourceExtractor, despite having a MARE of approximately 4\% as shown in Fig. \ref{fig: Flux_MARE}, tends to underestimate its \texttt{FLUXERR\_AUTO}. This results in a pronounced deviation from the ideal standard deviation value of 1 for the standardised residuals. 
In addition to previously discussed factors, the underperformance of SourceExtractor's \texttt{FLUXERR\_AUTO} in Fig. \ref{fig: Flux_WRSD} could also be attributed to its neglect of uncertainties associated with aperture scaling and centroid positioning.

\noindent
The ensemble method slightly underestimates the uncertainties across all S/N levels, albeit to a lesser extent than SourceExtractor. In contrast, ASID-FE's results more closely adhere to the ideal value, particularly at low and medium S/N levels, underscoring its superior calibration of uncertainties. It is worth noting that ASID-FE slightly overestimates the uncertainties at high S/N levels; however, its reliance on training with synthetic datasets offers an opportunity for refinement. A correction factor could be introduced to ensure that the estimated uncertainties and, consequentially, the standardised residuals align with a standard Gaussian distribution across all S/N levels. This potential for fine-tuning, coupled with the observed performance, underscores the robustness of the methods, with ASID-FE standing out for its efficiency and adaptability.

\noindent
In summary, the comparison between SourceExtractor and ASID-FE uncovers significant differences in error estimation. While SourceExtractor emphasises the variability induced by noise, ASID-FE incorporates potential errors in flux estimation. This comprehensive approach in ASID-FE enhances the understanding of the flux's true uncertainty and contributes to its superior performance in predicting astronomical sources.

\begin{figure*}[h!]

         \includegraphics[width=0.32\textwidth]{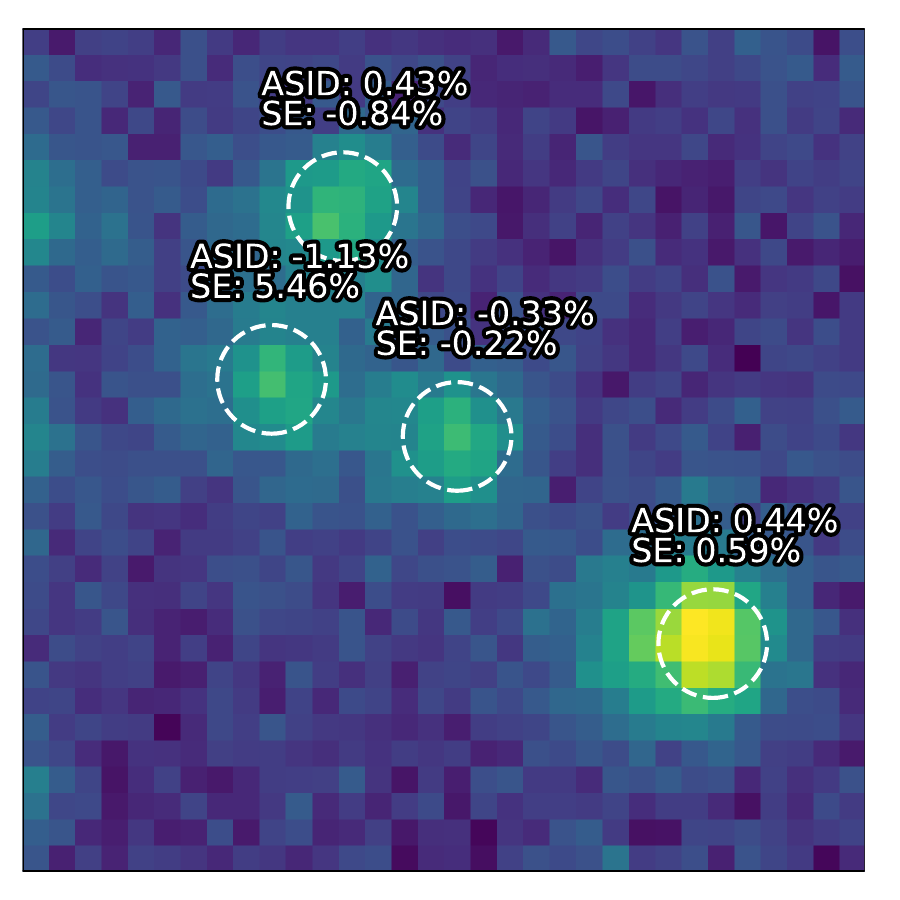}
         \hfill
         \includegraphics[width=0.32\textwidth]{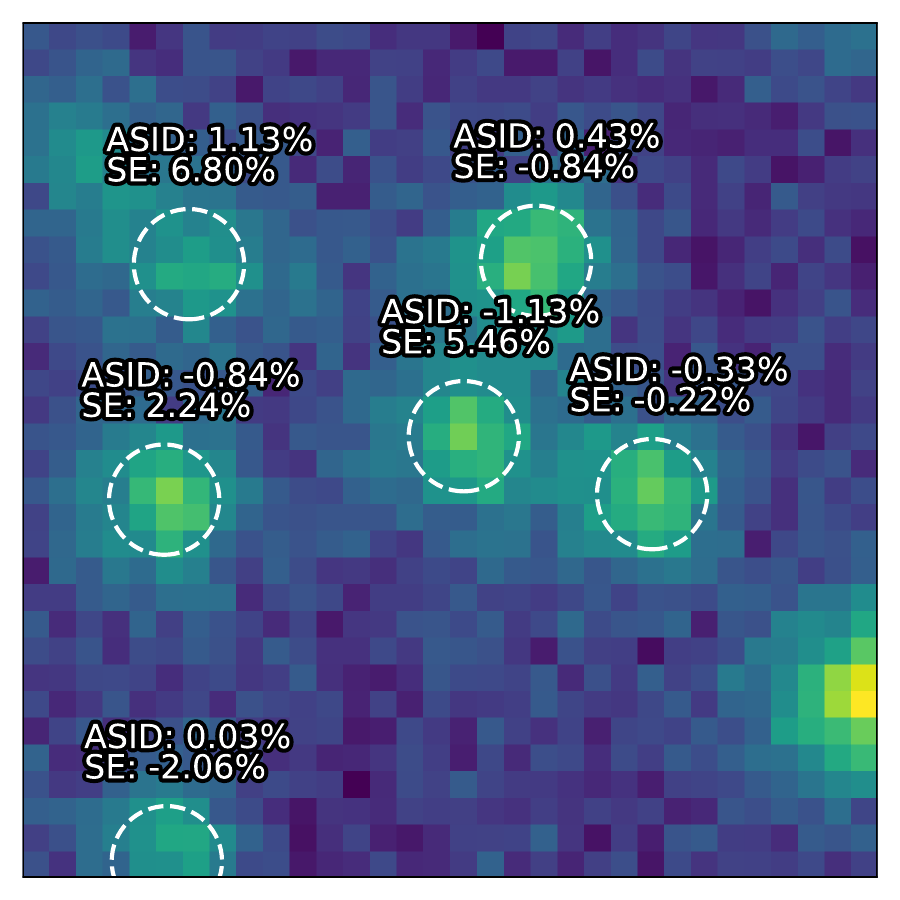}
         \hfill
         \includegraphics[width=0.32\textwidth]{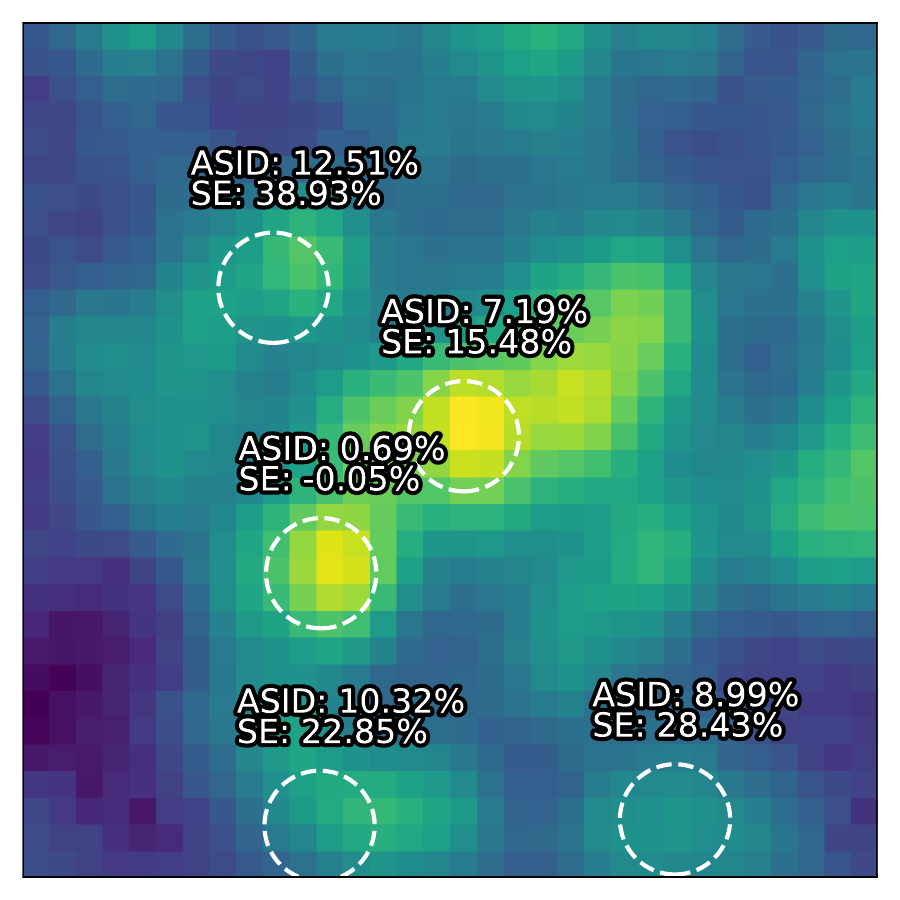}

\caption{Comparison of predicted flux percentage error for ASID-FE and SourceExtractor, highlighting their performance differences for different levels of image crowdedness.}
\label{fig: Crowded_images}
\end{figure*}

In addition to the quantitative metrics, in Fig. \ref{fig: Crowded_images}, we show the results obtained using both SourceExtractor and ASID-FE on three examples of crowded images. In all figures, the detected sources are indicated by the white circles, and the Percentage Error (PE) of the predicted fluxes for that source is written on top of each source. Only the sources detected by SourceExtractor are shown for comparison.

\noindent
The three images show that ASID-FE produces more accurate predictions than SourceExtractor, with fewer sources showing significant under- or over-predictions. The PE values for ASID-FE are generally smaller than those for SourceExtractor, indicating better overall performance in crowded regions. 
Overall, these visualisations provide a qualitative assessment of the performance of the two methods and highlight the superior performance of ASID-FE in predicting the properties of astronomical sources in crowded images.

\subsection{Real images application}
\label{sec: real application}

While synthetic datasets are needed for training models under controlled conditions, they cannot yet fully mimic the complexities of real-world data. Despite our best efforts to make them resemble real images, they lack the varied and intricate details found in actual MeerLICHT images. 
These details encompass factors such as atmospheric conditions, the telescope optics' unique characteristics, and the camera system's inherent noise. Hence, it becomes imperative to evaluate our model's performance within the context of real data, where these complexities are inherent.

For this study, we used the real dataset of MeerLICHT images described in Sec. \ref{sec: Real_dataset}. This dataset does not have ground truth labels, but we can still evaluate our model's performance by comparing its predictions to those made by established software such as SourceExtractor. This is a crucial step, as it helps us understand how well our model can generalise and how useful it might be in practical situations. Without real source properties in our dataset, we used two different approaches to test how well ASID-FE performs with real images. First, we applied the model, originally trained on synthetic images, directly to the real images. This helped us see how well ASID-FE could handle real data right off the bat. Then, we retrained the network from scratch, using predictions from SourceExtractor as training features. This allowed us to see how well ASID-FE could learn from and adapt to the complex features of real images.

To measure the similarity between predictions from our methods and those from SourceExtractor, we employed the Pearson correlation coefficient \citep{Pearson1895}, a statistical tool that quantifies the linear correlation between two data sets. More specifically, we compared the predictions made by ASID-FE from both the Direct and Scratch methodologies to those made by SourceExtractor. The results, as shown in Table \ref{table: pearson_synthetic}, reveal that the Pearson correlation coefficients for both the directly applied and the retrained ASID-FE models with respect to SourceExtractor are close to one. This implies a strong correlation, signifying that our models' predictions align closely with SourceExtractor's predictions.

\begin{table}
\caption{Comparison of Pearson correlation coefficients and ConTEST results for ASID-FE models applied to real MeerLICHT images. The Direct model was trained on synthetic images, while the Scratch model was retrained using SourceExtractor's predictions. Correlation coefficients assess the alignment of predicted $x$ and $y$ source positions and $flux$ values between ASID-FE and SourceExtractor. The ConTEST analysis evaluates the consistency between the two methods' predictions.}
\label{table: pearson_synthetic}      
\centering                          
\begin{tabular}{c c c c c c}        
\hline\hline                 
$Method$ & $x$ & $y$ & $flux$ & $Consistency$ \\    
\hline                        
Direct & 0.896 & 0.913 & 0.998 & Rejected\\      
Scratch & 0.956 & 0.952 & 0.999 & Not Rejected\\      

\hline                                   
\end{tabular}
\end{table}

Alongside correlation coefficients, we employed ConTEST \citep{Stoppa2023}, a robust nonparametric hypothesis testing method, to evaluate the consistency between the predictions yielded by SourceExtractor and ASID-FE. This statistical method allowed us to evaluate if the variations between the predictions, taking into account uncertainties, predominantly hover around zero. A non-rejection of the consistency hypothesis suggests a notable alignment between the predictions of the two methods.
However, applying ConTEST to the ASID-FE model trained on synthetic images and directly applied to real images led to the rejection of the consistency hypothesis for all three predicted quantities—x, y, and flux. This result points to discrepancies between this model's outputs and those generated by SourceExtractor.
On the other hand, when we applied ConTEST to the ASID-FE model that was retrained from scratch on real images using SourceExtractor's results, the consistency hypothesis was not rejected. This finding highlights the capacity of ASID-FE to adapt to real image data and produce results that align with established software such as SourceExtractor.

The ConTEST analysis revealed a significant divergence between the predictions of the ASID-FE model trained on synthetic images and then applied to real ones and the results from SourceExtractor. This difference underscores the inherent variation between synthetic and real images. While ASID-FE demonstrates a strong capability in predicting features when trained on them, the results suggest that the main determinant is the nature of the training images themselves.
In future studies, we aim to address this discrepancy between synthetic and real images more effectively. By integrating software such as Pyxel \citep{Arko2022}, we hope to train on images that more closely resemble real-world scenarios, minimising the potential performance drop when a model trained on synthetic images is used for prediction on real-world data. Ultimately, these advancements will serve to close the reality gap, enhancing the model's applicability and reliability in real-world applications \citep{Caron2022}.

\subsection{Transfer learning on ZTF images}
\label{sec: transfer learning}

There are multiple approaches to transferring the knowledge of a network to a similar application. In their study, \cite{Yosinski2014} found that deep neural networks tend to learn a hierarchy of features, where the first layers capture more general, low-level features (e.g. edges and textures), while the last layers capture more task-specific, high-level features (e.g. object parts and shapes). This finding gives excellent insight into the use of transfer learning in fine-tuning models for specific tasks. Additional studies have also tried quantifying this behaviour \citep{Orhand2021}.

In this section, we discuss the application of transfer learning on our regression model trained on MeerLICHT synthetic data and its application to real images from the Zwicky Transient Facility (ZTF, \citealp{Bellm2019}). ZTF is a large-scale astronomical survey designed to study the dynamic sky in the optical regime.
However, there are notable differences between the MeerLICHT and ZTF images, most prominently concerning spatial resolution and sky coverage. ZTF provides a larger field of view with lower resolution images than MeerLICHT, impacting the detectability of faint sources and the precision of extracted features. Further, variations may arise from different filters used by the two surveys, affecting the observed fluxes of astronomical sources.

\noindent
As for Sec. \ref{sec: real application}, due to the lack of exact features in real images, we use DAOPHOT \citep{Stetson1987} catalogues, which are paired with ZTF images, as a benchmark. DAOPHOT is a renowned software in the astronomical community, offering tools for detecting, measuring, and analysing the properties of point sources in astronomical images. Comparing our transfer learning model's results with the DAOPHOT catalogues will shed light on our methodology's real-world effectiveness and highlight areas requiring improvement.

We adopt three distinct strategies to evaluate the model, initially trained on synthetic MeerLICHT images on ZTF data. First, we directly apply the MeerLICHT-trained model. Next, considering the significant differences between the telescopes, we fine-tune the model in two unique ways. In the first approach, we retain the weights of the early layers and retrain the final layers using real ZTF data and Daophot's results. In the second, we use the full pre-trained model on synthetic images as a baseline, retraining it entirely with real ZTF data. Additionally, we also explore a ground-up approach, where we train a new network from scratch using DAOPHOT's predictions as training data. We name the four applications Direct, Frozen, Retrained and Scratch; the Pearson coefficient for all the methods is calculated and presented in Table \ref{table: pearson_ZTF}.

\begin{table}

\caption{Pearson correlation coefficient and consistency test between ASID-FE and DAOPHOT for the four different transfer learning methods.}
\label{table: pearson_ZTF}      
\centering                          
\begin{tabular}{c c c c c}        
\hline\hline                 
$Method$ & $x$ & $y$ & $flux$ & Consistency \\    
\hline                        
Direct & 0.723 & 0.707 & 0.982 & Rejected \\      
Frozen & 0.833 & 0.803 & 0.998 & Rejected \\
Retrained & 0.938 & 0.930 & 0.999 & Not Rejected \\      
Scratch & 0.958 & 0.954 & 0.999 & Not Rejected \\

\hline                                   
\end{tabular}
\end{table}

Our investigation revealed that both the Direct application and the Frozen approach, where the early layers were kept constant, and the rest were allowed to learn, failed to produce optimal outcomes. The suboptimal performance can be largely attributed to the differences in the PSF size - a crucial low-level feature that varies between MeerLICHT and ZTF images. As such, preserving the early layers of the architecture, which are fundamental for learning this feature, turned out to be counterproductive.

\noindent
The Retrained approach, rooted in the synthetic MeerLICHT images, yielded promising results. It showed performance on par with the Scratch model trained from real ZTF data. Both these methods showcased outcomes that aligned well with the benchmarks set by the official DAOPHOT catalogues, providing an encouraging indication of our methodology's potential effectiveness. To further validate these findings, we deployed the ConTEST statistical test, introduced in the previous section, to assess the consistency against DAOPHOT predictions. As anticipated, the null hypothesis of consistency is not rejected only for the Retrained and Scratch methods, which further endorses their superior performance.

Through this investigation, it becomes clear that while a model trained on synthetic data from one telescope can technically be applied to data from a different telescope, the process often necessitates significant adjustments or comprehensive retraining to cater to the unique characteristics of real data derived from different telescopes. This is especially the case when there is a substantial difference in the angular resolution between telescopes.
Given these findings, we recommend against the practice of transfer learning in such scenarios. If a synthetic dataset tailored specifically to the target telescope (such as ZTF) is available, training a model directly on this synthetic dataset is the superior approach. This highlights the critical importance of generating and employing synthetic datasets designed to match the specific characteristics of the telescopes used in the study.

\section{Discussion on PSF}

For many years, the point spread function has been an essential tool for astronomers to estimate the flux of sources in astronomical images accurately. The PSF describes the spread of light from a point source in the image, and its knowledge helps to separate sources that are close together and deconvolve the PSF effects from the measured flux.

However, traditional methods for flux estimation relying on knowledge of the PSF have several limitations. In crowded regions with closely spaced sources, it can be challenging to accurately separate the sources and estimate their fluxes, even with knowledge of the PSF. Furthermore, for observations with varying PSFs across the image, PSF information may not be readily available or accurate, leading to inaccuracies in flux estimation.
This paper explores the possibility of estimating the flux of astronomical sources without explicit knowledge of the PSF. We leverage regression algorithms, a subset of machine learning techniques, to model the inherent distribution of sources in an image to predict their fluxes. These algorithms offer a complementary approach to traditional methods, demonstrating the ability to adapt and provide precise flux estimates across various astronomical datasets.

It is essential to consider, however, that while machine learning offers considerable benefits and possibilities, its effectiveness depends on the training data's quality and the chosen model's appropriateness. Despite these advancements, the value of PSF information in flux estimation should not be undermined, as it still serves a critical role in numerous scenarios.

\section{Conclusions}

By implementing a two-step mean variance estimation network, we have created a novel method for estimating features of astronomical sources in images, known as ASID-FE. This technique involves a two-fold process: initially, the network estimates the source's centre coordinates and flux through a convolutional neural network trained on synthetic images. The second stage harnesses the insights gained from the first phase to fine-tune the predictions of x, y, and flux while simultaneously characterising the uncertainties tied to these predictions.

Through rigorous testing on both synthetic and real images, we have verified the robustness and precision of our method in estimating the properties of astronomical sources. The unique two-step process of our methodology allows it to outperform simple MVE networks. Our evaluation showed that when dealing with synthetic images with known true values, ASID-FE results show less bias than those yielded by SourceExtractor. Furthermore, our technique demonstrates superior proficiency in characterising uncertainties. Unlike traditional methods, ASID-FE estimates a more detailed and nuanced uncertainty that often goes overlooked, providing a richer understanding of the underlying statistical properties.

In summary, ASID-FE presents an efficient, versatile tool for the estimation of properties of astronomical sources. Our method holds distinct advantages over traditional methodologies, primarily its capacity to learn from large synthetic datasets with known ground truth and effectively apply these insights to real images. Looking forward, we aim to explore methods to minimise further the disparity between synthetic and real images, including the potential integration of software such as Pyxel and ScopeSim. By doing so, we hope to ensure a minimal loss in performance when making predictions using models trained on synthetic images. As we continue refining our methodology, we are optimistic about improving the accuracy and versatility of ASID-FE across various astronomical contexts.

The straightforward nature of our introduced TS-MVE method lends itself to the substantial potential for broader applications, particularly concerning feature regression tasks in images. Future avenues of exploration could include assessing our method's generalisability across diverse astronomical datasets and exploring its potential for multi-wavelength analysis of astronomical sources.

\begin{acknowledgements}
F.S. and G.N. acknowledge support from the Dutch Science Foundation NWO. S.B. and G.Z. acknowledge the financial support from the Slovenian Research Agency (grants P1-0031, I0-0033 and J1-1700). R.R. acknowledges support from the Ministerio de Ciencia e Innovación (PID2020-113644GB-I00). G.P. acknowledges support by ICSC – Centro Nazionale di Ricerca in High Performance Computing, Big Data and Quantum Computing, funded by European Union – NextGenerationEU.
P.J.G. is partly supported by NRF SARChI Grant 111692.
The MeerLICHT telescope is a collaboration between Radboud University, the University of Cape Town, the South African Astronomical Observatory, the University of Oxford, the University of Manchester and the University of Amsterdam, and supported by the NWO and NRF Funding agencies. 
The authors would like to thank Dr Dmitry Malyshev for his valuable suggestions and insightful discussions, which contributed to improving this paper.
\end{acknowledgements}

\bibliographystyle{aa}
\bibliography{Bibliography}

\end{document}